\documentclass[12pt,preprint]{aastex61}
\usepackage{graphicx}

\makeatletter
\newcommand\footnoteref[1]{\protected@xdef\@thefnmark{\ref{#1}}\@footnotemark}
\makeatother

\begin{document}

\shortauthors{}
\shorttitle{}

\title{IRAS 09002-4732: A Laboratory for the Formation of Rich Stellar Clusters}

\author{Konstantin V. Getman}
\affiliation{Department of Astronomy \& Astrophysics, Pennsylvania State University, 525 Davey Lab, University Park, PA 16802}

\author{Eric D. Feigelson}
\affiliation{Department of Astronomy \& Astrophysics, Pennsylvania State University, 525 Davey Lab, University Park, PA 16802}
\affiliation{Center for Exoplanets and Habitable Worlds, Pennsylvania State University}

\author{Michael A. Kuhn}
\affiliation{Department of Astronomy, California Institute of Technology, Pasadena CA  91125}

\author{Patrick S. Broos}
\affiliation{Department of Astronomy \& Astrophysics, Pennsylvania State University, 525 Davey Lab, University Park, PA 16802}

\author{Gordon P. Garmire}
\affiliation{Department of Astronomy \& Astrophysics, Pennsylvania State University, 525 Davey Lab, University Park, PA 16802}

\correspondingauthor{Konstantin V. Getman}
\email{gkosta@astro.psu.edu}

\begin{abstract}
IRAS 09002-4732 is a poorly studied embedded cluster of stars in the Vela Molecular Ridge at a distance of 1.7~kpc.  Deep observations with the {\it Chandra} X-ray Observatory, combined with existing optical and infrared surveys, produce a catalog of 441 probable pre-main sequence members of the region.  The stellar spatial distribution has two components: most stars reside in a rich, compact, elliptical cluster, but a minority reside within a molecular filament several parsecs long that straddles the cluster.  The filament has active distributed star formation with dozens of unclustered protostars.  The cluster pre-main sequence population is $\leq 0.8$~Myr old and deeply embedded; its most massive member is extremely young producing an ultracompact H~II region.  The cluster total population deduced from the X-ray luminosity function is surprisingly rich, twice that of the Orion Nebula Cluster.  The cluster core is remarkably dense where strong N-body interactions should be occurring; its Initial Mass Function may be deficient in massive stars.  We infer that IRAS 09002-4732 is a rare case where a rich cluster is forming today in a molecular filament, consistent with astrophysical models of cluster formation in clouds that involve the hierarchical formation and merging of groups in molecular filaments.
\end{abstract}


\section{Introduction \label{intro.sec}}

\subsection{The formation of star clusters} \label{form.sec}

Most stars in the Galaxy are field stars but originate in rich clusters within giant molecular cloud complexes \citep{Lada03}.  The mechanisms by which rich clusters form has not been easy to discern.  An understanding is emerging that rich clusters do not form monolithically in a single event of gravitational collapse, but rather are produced hierarchically from the merger of small groups that collapse asynchronously in molecular filaments within cloud complexes.  Hydrodynamical calculations of star formation under realistic cloud conditions suggest that this hierarchical cluster formation process extending over millions of years, slowed by turbulence and stellar feedback \citep{Bate03, Federrath15, Vazquez17, Vazquez19}.  Empirical evidence for this picture has been accumulating for this scenario from disparate findings: the ubiquity of turbulence in molecular clouds \citep{MacLow04}, the prevalence of filamentary molecular structures \citep{Andre10}, detection of complicated kinematics in filaments including gravitational infall  \citep{Schneider10, Zhang17}, and discovery of spatio-temporal gradients in rich young clusters \citep{Getman18}.  The latter phenomenon, unexpected from simple cluster formation models, requires continuing star formation in cluster cores. 

But a clear example of a rich cluster $-$ that is, thousands of stars dominated by massive OB stars $-$ in the solar vicinity that is now forming from infalling and merging filaments has been difficult to find.  (A young cluster with 1000 members will typically have a star with maximum mass $\sim 40$~M$_\odot$; Popescu \& Hanson~2014.)  The Spokes Cluster in the NGC 2264 star forming region at $d \simeq 0.8$~kpc has over a dozen protostars closely associated with dusty filaments, but the dominant star does not form an H~II region, the cluster population is only a few hundred stars,  and the filaments are $< 1$~pc in length \citep{Teixeira06, Kuhn15}.    IC 5146 (= Cocoon Nebula) with $d \simeq 0.8$~kpc is an embedded cluster with young protostars, dominated by a B0 star with several short ($<1$~pc) dusty filaments; but its total population does not exceed a couple of hundred stars \citep{Johnstone17}.   Many more ultracompact H~II regions tracing current OB star formation are known at distances $3-15$~kpc where the greater distance inhibits detailed study \citep{Wood89}.  Perhaps the most interesting case is W3~Main at $d \simeq 2$~kpc that is a populous, spherical, embedded cluster with at least $\sim 2000$ stars and a remarkable concentration of OB stars of different ages in the cluster core \citep{Tieftrunk97, Feigelson08, Broos13, Bik14}.  A small molecular filament ($0.5 \times <0.1$~pc) has been identified among the W3~Main H~II regions and the cluster appears to be fed by several larger dusty filaments several parsecs in length \citep{Tieftrunk98, Rivera13}.  

The most detailed observational examination of star formation in massive clouds has been made in the nearby the Orion molecular cloud complex at distance $d \simeq 0.4$ kpc \citep{Bally08, Rezaei18}.  It shows that the process of large-scale star formation can be very complicated, extending over tens of parsecs and millions of years.  

Current active star formation in the filamentary Orion A cloud is concentrated along its head with projected length $\simeq 10$~pc, such as the protostars are distributed in the OMC 2/3 region, and in the small clusters associated with the Becklin-Neugebauer object in OMC-1 \citep{Johnstone99, Grosso05, Megeath12}.  The present-day molecular cloud is surrounded by thousands of stars, some associated with distinct OB associations and others probably drifted from the current star forming filament \citep{Blaauw91, Carpenter01, Bouy14, Kounkel18}.  Dozens of supernova explosions produced hot bubbles spanning $>100$ pc.  Smaller clusters formed a few million years ago are still present as coherent structures with proper motions reflecting the motions of cloud components that are now partially or completely dissipated \citep{Getman19}.   

The most prominent structure is the Orion Nebula Cluster (ONC), a compact rich cluster of several thousand stars, mostly $1-3$ Myr old, with a dense central concentration of higher-mass stars known as the Trapezium \citep{Hillenbrand97, Odell01}.  However, the molecular cloud structure giving rise to the ONC is now dissipated, so the formation process of the ONC is uncertain.

\subsection{IRAS 09002-4732: A Laboratory for Rich Cluster Formation} \label{whyI.sec}

We discuss here IRAS 09002-4732, a young stellar cluster that resembles the ONC $-$ both are associated with a molecular filament several parsecs long $-$  but at an earlier stage of evolution. The principal infrared and molecular studies of the region are by \citet{Lenzen91} and \citet{Apai05}.  Located in the Vela Molecular Ridge (VMR-A), the stellar population is concentrated in a cluster still embedded in, and likely destroying, its natal cloud.  Its dominant star has spectral type around O7 similar to $\theta^1$C~Ori in the ONC but is heavily obscured with $A_V \simeq 20$~mag, producing a cometary ultracompact H~II region $\simeq 0.05$~pc in extent.  In contrast,  $\theta^1$C~Ori  and its nearby Trapezium stars are unobscured and produce a giant H~II region $>2$~pc in extent.   In both cases, the dominant star is projected close to dense molecular cores \citep{Lapinov98} although in the ONC case the star cluster lies in the foreground evacuated H~II region.   The molecular properties and massive stellar product of the nearby VMR-C molecular region is also similar to the Orion A cloud \citep{Yamaguchi99}.   
 
\begin{figure}[ht]
\centering
\includegraphics[width=1.0\textwidth]{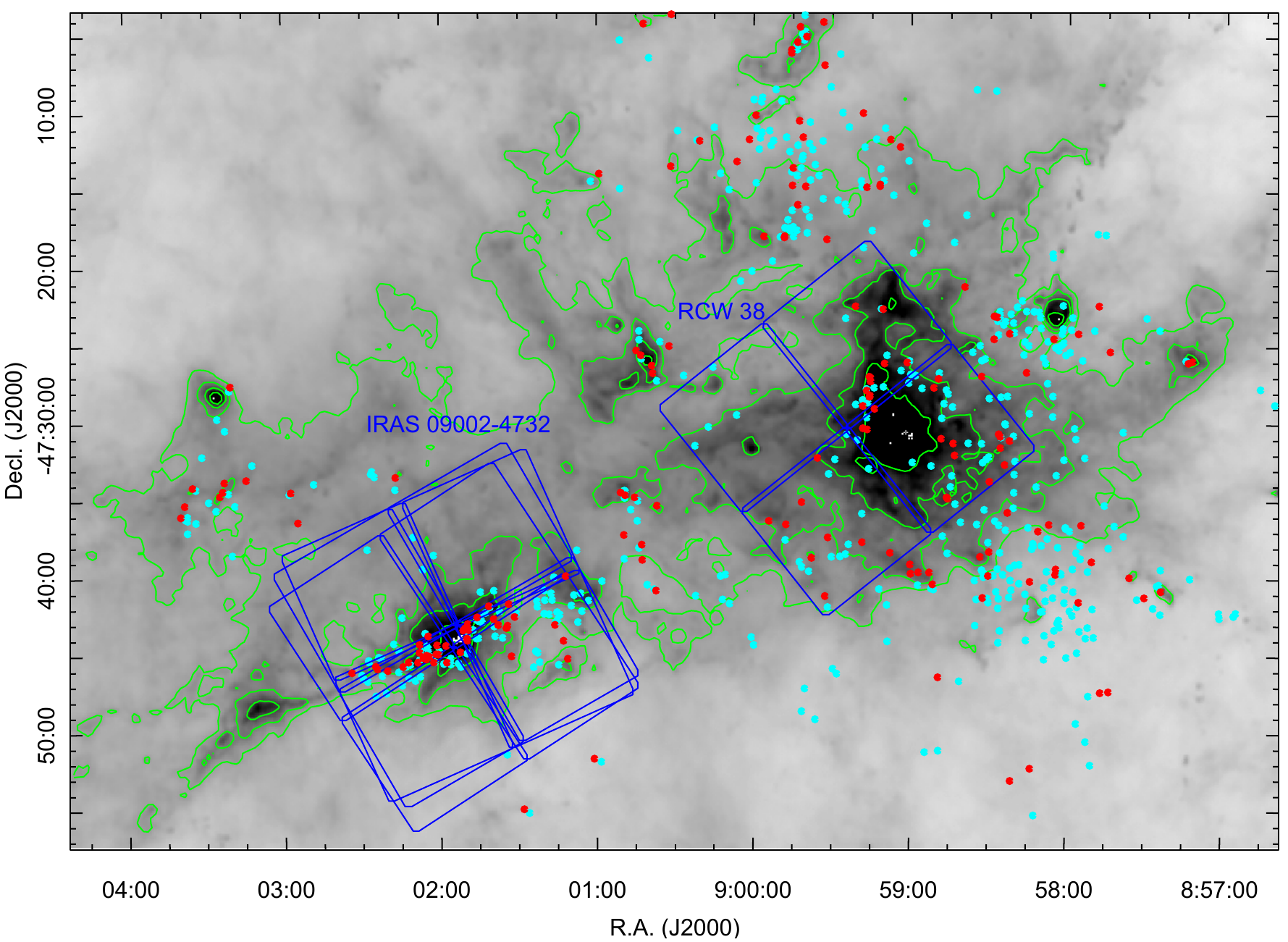}
\caption{A $1.3^\circ \times 0.9^\circ$ ($39 \times 27$~pc at a distance of 1.7~kpc) dust map from the 250~$\mu$m  band Herschel SPIRE instrument showing the IRAS 09002-4732 and RCW 38 star forming regions with their environs. Green contours show levels of infrared surface brightness.   Circles show Spitzer-selected infrared-excess stars: red = Stage 1 corresponding to Class 0/I protostars, and cyan = Stage 2 corresponding to Class II/III pre-main sequence stars \citep{Povich13}.  The blue polygons show the MYStIX and IRAS 09002-4732  {\it Chandra} fields.  \label{Herschel-Chandra.fig}}
\end{figure}

We show in this study that the analogy between the ONC and IRAS 09002-4732 is strong by improving the stellar census of the IRAS 09002-4732 cloud, comparing its properties to other nearby rich young clusters, and demonstrating that it lies within a molecular filament populated by many protostars.  Our group encountered this active star formation in the larger elongated structure while constructing the MYStIX Infrared-Excess Source \citep[MIRES;][]{Povich13} catalog members associated with the nearby RCW~38 rich cluster.  Figure~\ref{Herschel-Chandra.fig} shows a wide area of the infrared sky encompassing the RCW 38 and IRAS 09002-4732 regions with MIRES selected stars superimposed.

Our follow-up study found a rich group of 24 candidate Class 0/I protostars in the region $09^h 01^m 12^s < \alpha < 09^\circ 02^\prime 24^{\prime\prime}$ and -47:40$< \delta < $-47:45 \citep{Romine16}.  They lie along a dusty filament seen in $Herschel$ infrared maps that is oriented northwest to southeast, bifurcated by the older IRAS 09002-4732 star cluster.  Romine et al.\ selected the protostars based on UKIDSS $JHK$ and Spitzer 3.6-8.0~$\mu$m spectral energy distributions; several also exhibit a 4.5~$\mu$m excess characteristic of shocked molecular protostellar outflows \citep{Cyganowski08}.  These photometrically selected protostars have a wide range of luminosities with 8~$\mu$m brightnesses ranging from 5 to 12 mag.  Some are quite massive protostars, but the brightest is two orders of magnitude less luminous than the dominant O7 star of the IRAS 09002-4732 cluster (\S\ref{core.sec}).  

\citet{Povich13} and \citet{Romine16} thus found active star formation in a molecular filament around the somewhat older (and much richer) IRAS 09002-4732 star cluster, all of which appears younger than the ONC.   This motivates further study to characterize the IRAS 09002-4732 region more completely, compare it to other nearby young clusters, and investigate whether empirical insights can emerge view on the birth of ONC-type rich clusters.

\subsection{The Role of X-ray Imaging} \label{whyX.sec}

Our observational effort here starts with an improvement in the stellar census of IRAS 09002-4732, both in its central cluster and distributed in the surrounding molecular material.  This is achieved with deep exposures using NASA's {\it Chandra} X-ray Observatory, together with complementary infrared surveys.  {\it Chandra} effectively detects rich pre-main sequence (PMS) populations in star forming regions out to several kiloparsecs in the Galactic Plane primarily due to enhanced magnetic dynamos, and subsequent magnetic reconnection flaring, in fully convective rapidly rotating stars \citep[see reviews by][]{Gudel07, Feigelson18}.  Solar-mass PMS stars are typically $10^2-10^3$ times more X-ray luminous that the contemporary Sun, and can thus be readily differentiated from most main sequence field stars by virtue of elevated X-ray emission.  The flare X-ray spectrum is harder than typically seen in solar flares, so PMS stars can often be detected even if embedded in their natal cloud. 

Our group has produced catalogs of $>$40,000 PMS stars with X-ray and infrared surveys in $\simeq 40$ OB-dominated clusters in massive star forming complexes out to distances $\simeq 3$~kpc \citep{Getman05, Townsley11a, Feigelson13, Broos13, Getman17}.  In addition to providing an improved census of PMS members of a star forming region, the X-ray photometry and spectra, together with near-infrared photometry, allow estimates of individual stellar masses, ages, and line-of-sight absorptions \citep{Getman10, Getman14}.   We also use X-ray properties as measures of the spatial structure and total population of the PMS populations \citep{Kuhn14, Kuhn15}.   With this multifaceted characterization of the stellar population produced in a molecular cloud over millions of years, new empirically-based insights into the cluster formation process can be sought.  Recent results from some of these X-ray studies are reviewed by \citet{Feigelson18}.  

The X-ray observations of IRAS 09002-4732, together with methods of data reduction and inference of astronomical properties are described in \S\ref{datameth.sec}.  The results of {\it Chandra} X-ray imaging and catalog of probable cloud members are presented in \S\ref{members.sec}.  Inference of stellar and cluster properties $-$ initial mass function, total population, age, spatial structure and stellar densities $-$ are developed in \S\ref{properties.sec}.  The final section (\S\ref{discussion.sec}) discusses the star formation history of the region, and its role as a young analog of the ONC region.  Appendices provide details on the stars in the core of the IRAS 09002-4732 cluster and stellar ages in the nearby rich RCW~38 cluster.  

\section{Data and Analysis} \label{datameth.sec}

Three observations of IRAS 09002-4732 were obtained during {\it Chandra} X-ray Obervatory Cycle 18 using the Advanced CCD Imaging Spectrometer imaging array \citep[ACIS-I;][]{Garmire03}.  The $17^\prime \times 17^\prime$ image from four contiguous CCD chips was centered at ($\alpha, \delta$)$_{J2000}$ = (09:01:55.00, -47:43:46.8) and the total exposure was close to 100~ks.   The instrument was operated in Very Faint Timed Exposure mode.  Exposure details are provided in Table~\ref{CXO.tbl}.  Data quality is excellent.  

Data analysis closely followed procedures from our earlier projects studying PMS populations (references above), and is briefly summarized here.  The suite of  {\it ACIS Extract} procedures described by \citet{Broos10} provides very sensitive and reliable identification of faint X-ray sources in {\it Chandra} images \footnote{
Some more elaborate methods used in some of our previous studies $-$ such as probabilistic matching to infrared catalogs \citep{Naylor13},  parametric modeling field star and extragalactic contaminant populations, and naive Bayes classification to reduce contamination \citep{Broos13} $-$ were not applied here.
}.

\begin{deluxetable}{lllc}
\tablecaption{Chandra Exposures \label{CXO.tbl}}
\tabletypesize{\footnotesize}
\tablewidth{0pt}
\centering
\tablehead{
\colhead{ObsID} & \colhead{Type\tablenotemark{a}} & \colhead{Date} & \colhead{Exp\tablenotemark{b} (ks)} 
}
\startdata
18891 & GO   & 2017-03-03 & 49.3 \\
19710 & GTO & 2017-09-04 & 28.3 \\
20727 & GTO & 2017-09-09 & 21.8 \\
\enddata
\tablenotetext{a}{GO = Guest Observer;  GTO = Guaranteed Time Observation, {\it Chandra} ACIS Team} 
\tablenotetext{b}{Exposure times are the net usable times after removal of high instrumental background}
\end{deluxetable}

\begin{enumerate}

\item A deep catalog of candidate sources is obtained from bumps in a smoothed map of the field based on maximum likelihood deconvolution using the known spatially-variable instrument point spread function.  Local background levels are iteratively calculated for each candidate source; this is needed because the three exposures have misaligned roll angles and chip gaps.  This candidate source identification procedure is performed on the merged image from the three exposures in Table~\ref{CXO.tbl}.   

\item A global astrometric correction to the pointing direction is made by removing any offset between bright X-ray sources and 2MASS counterparts.  This allows sub-arcsecond positional accuracy for on-axis sources. 

\item  Photons are extracted for each candidate source in small regions scaled to the local point spread function, typically containing 90\% of the expected photons.  Local background rates are identified and subtracted to give net source counts.  The local exposure time is obtained from the merged exposure map.  

\item A variety of X-ray properties are then calculated from the extracted events: net count rate corrected for local exposure time and point spread function tails;  source locations with errors depending on off-axis angle and net count rate; probability of source existence based on Poisson distributions for the source and background photon rates; hypothesis tests for variability in the photon arrival times using a Kolmogorov-Smirnov test; and median energy of the net counts.  

\item Line-of-sight interstellar extinction to the source is obtained from the median photon energy and an absorption-corrected X-ray luminosity is obtained for each source.  These calculations, based on the {\it XPHOT} method described by \citet{Getman10}, are restricted to sources with $>5-7$ net counts.  
\end{enumerate}

\vfill 

\begin{deluxetable}{crrrrrrrrrccc}
\tabletypesize{\scriptsize} 
\tablecaption{Candidate X-ray Sources \label{candx.tbl}}
\tablehead{
   & \multicolumn{4}{c}{Position} & \multicolumn{5}{c}{Extraction} & \multicolumn{3}{c}{Characteristics} \\ [-8pt]                  
    & \multicolumn{4}{c}{\hrulefill} &  \multicolumn{5}{c}{\hrulefill} & \multicolumn{3}{c}{\hrulefill} \\ 
\colhead{CXOU J} & \colhead{$\alpha$}  & \colhead{$\delta$} & \colhead{Error} & \colhead{$\theta$} &
\colhead{$C_{net}$} & \colhead{$\sigma_{net}$} & \colhead{$B$} & \colhead{PSF} &  \colhead{$\log P_B$} & 
\colhead{Anom} & \colhead{Var}  & \colhead{$ME$}  \\ [-4pt]
\nocolhead{} & \multicolumn{2}{c}{J2000} & \colhead{\arcsec} & \colhead{\arcmin} & 
\colhead{counts} & \colhead{counts} & \colhead{counts} & \colhead{Frac} & \colhead{} &
\colhead{} & \colhead{} &  \colhead{keV}
}
\startdata
090106.38$-$474137.6 &135.276618 & -47.693802 &    1.26 &     8.0 &    4.8 &     3.0 &     4.2 &    0.90 &    -2.9 & \nodata &  \nodata &     2.9   \\ %
090106.52$-$474429.7 &135.277185 & -47.741607 &    0.69 &     8.2 &   30.3 &     6.4 &     9.7 &    0.90 &   $<$-5 & \nodata &        2 &     2.6   \\ %
090107.22$-$473954.1 &135.280123 & -47.665039 &    2.26 &     8.5 &    1.6 &     1.7 &     1.4 &    0.61 &    -1.1 & g       &  \nodata &     1.5   \\ %
090107.84$-$474057.9 &135.282690 & -47.682762 &    1.10 &     8.0 &    4.8 &     2.5 &     1.2 &    0.56 &    -2.9 & \nodata &        1 &     1.7   \\ %
090109.07$-$474108.4 &135.287798 & -47.685693 &    1.16 &     8.1 &    6.6 &     3.9 &     8.4 &    0.90 &    -2.2 & g       &  \nodata &     5.4   \\ %
090109.31$-$474708.3 &135.288823 & -47.785649 &    1.93 &     8.4 &    2.3 &     2.8 &    10.7 &    0.90 &    -0.9 & \nodata &  \nodata &     3.9   \\ %
090109.56$-$474426.0 &135.289860 & -47.740581 &    0.96 &     7.7 &    9.4 &     4.2 &     7.6 &    0.90 &    -4.0 & \nodata &        1 &     2.1   \\ %
090109.84$-$474149.0 &135.291008 & -47.696972 &    0.56 &     7.8 &   28.5 &     6.2 &     8.5 &    0.90 &   $<$-5 & \nodata &  \nodata &     2.3   \\ %
090144.19$-$474000.9 &135.434129 & -47.666930 &    0.06 &     4.2 &  287.8 &    17.1 &     1.2 &    0.90 &   $<$-5 & \nodata &        2 &     1.3   \\ %
090146.08$-$474957.9 &135.442019 & -47.832768 &    0.15 &     6.4 &  191.8 &    14.1 &     4.2 &    0.90 &   $<$-5 & \nodata &        2 &     2.6   \\ %
090154.63$-$474411.1 &135.477666 & -47.736417 &    0.19 &     0.6 &    5.2 &     2.5 &     0.8 &    0.90 &    -3.8 & g       &  \nodata &     5.0   \\ %
090201.21$-$474455.9 &135.505080 & -47.748875 &    0.27 &     1.6 &    2.8 &     1.8 &     0.2 &    0.90 &    -4.1 & g       &  \nodata &     3.1   \\ %
090153.81$-$474415.7 &135.474218 & -47.737710 &    0.10 &     0.6 &   17.6 &     4.3 &     0.4 &    0.91 &   $<$-5 & ga      &  \nodata &     2.3   \\ %
090154.10$-$474412.1 &135.475448 & -47.736718 &    0.20 &     0.6 &    4.3 &     2.3 &     0.7 &    0.90 &    -2.8 & g       &  \nodata &     4.0   \\ %
090154.11$-$474410.3 &135.475484 & -47.736197 &    0.10 &     0.5 &   19.9 &     4.6 &     1.1 &    0.90 &   $<$-5 & g       &  \nodata &     3.1   \\ %
090154.29$-$474408.8 &135.476224 & -47.735789 &    0.11 &     0.5 &    5.8 &     2.9 &     2.2 &    0.55 &    -2.8 & g       &  \nodata &     4.8   \\ %
090154.33$-$474410.3 &135.476415 & -47.736201 &    0.03 &     0.5 &  111.1 &    10.7 &     1.9 &    0.75 &  $<$-5 & g       &  \nodata &     3.8   \\ %
090154.36$-$474408.5 &135.476522 & -47.735720 &    0.10 &     0.5 &    6.9 &     2.9 &     1.1 &    0.57 &    -4.5 & g       &  \nodata &     4.1   \\ %
090154.38$-$474409.5 &135.476588 & -47.735981 &    0.09 &     0.5 &    4.3 &     2.7 &     2.7 &    0.42 &    -3.8 & g       &  \nodata &     5.1   \\ %
090154.63$-$474411.1 &135.477666 & -47.736417 &    0.19 &     0.6 &    5.2 &     2.5 &     0.8 &    0.90 &    -3.8 & g       &  \nodata &     5.0   \\ %
090154.75$-$474421.6 &135.478157 & -47.739350 &    0.23 &     0.7 &    2.7 &     1.8 &     0.3 &    0.86 &    -2.5 & g       &  \nodata &     3.9   \\ %
090154.78$-$474407.7 &135.478276 & -47.735498 &    0.21 &     0.5 &    3.6 &     2.1 &     0.4 &    0.83 &    -3.4 & \nodata &  \nodata &     3.5   \\ %
090154.89$-$474406.7 &135.478713 & -47.735210 &    0.23 &     0.5 &    2.6 &     1.8 &     0.4 &    0.81 &    -2.3 & \nodata &  \nodata &     5.4   \\ %
090155.32$-$474412.9 &135.480536 & -47.736937 &    0.26 &     0.6 &    2.6 &     1.8 &     0.4 &    0.91 &    -2.2 & g       &  \nodata &     4.2   \\ %
\enddata
\tablecomments{
This table with 1034 lines is available in its entirety in the electronic edition of the Journal.  A few sources are shown here for guidance regarding its form and content. Column definitions: \vspace{3pt} \\
CXOU:  {\it Chandra} X-ray Observatory Unidentified source  \vspace{3pt} \\ 
$\alpha$, $\delta$, Err:  Right ascension and declination (in decimal degrees) for epoch J2000.0 with the 63\% error circle representing the random component of position uncertainty.  The field is astrometrically aligned to the 2MASS reference frame. \vspace{3pt} \\
$\theta$: Off-axis angle (in arc minutes) from the pointing direction of the ACIS imager.  The mirror point spread functions deteriorates nonlinearly as $\theta$ increases. \vspace{3pt} \\
$C_{net}$, $\sigma_{net}$, $B$, PSF Frac, $\log P_B$:  Counts extracted in the PSF-shaped polygon for source characterization in the $0.5-8$~keV band.  The total number of extracted counts is $C_{net} + B$ where $B$ is the local background counts scaled to the extraction area and $C_{net}$ is the net source counts after background subtraction.  PSF Frac is the fraction of the local point spread function captured in the extraction polygon. A reduced PSF fraction (significantly below 90\%) may indicate that the source is in a crowded region. $\log P_B$ is the logarithm of the probability that the source is spurious (specifically, that the observed total number of extracted counts will occur in a single trial of the Poisson distribution given the observed background counts $B$).    \vspace{3pt} \\
Anom: Anomaly flag.  ``a'' reports that photometry and spectrum may contain $>10$\% cosmic ray afterglow events in the detector.  ``g''  reports that the source spent more than 10\% of the observation on dead regions of the detector (principally chip gaps). \vspace{3pt} \\
Var: Variability flag based on the Kolmogorov-Smirnov test.  ``...'' indicates no evidence of variability.  ``1'' indicates possible ($0.005 < P_{KS} < 0.05$) variability.  ``2'' indicates very probable variability ($P_{KS} < 0.005$).  Variability can be present within a single {\it Chandra} exposure, in the merged {\it Chandra} exposures, or between averaged {\it Chandra} exposures. Spurious indications can be produced by satellite dithering across dead regions of the detector, or by variations in the background rather than source photon arrival rate.   \vspace{3pt} \\
ME:  Median energy (in keV) of the background-subtracted photons.
}
\end{deluxetable}

\begin{deluxetable}{ccccccccccc}
\centering
\tabletypesize{\scriptsize} 
\tablecaption{X-ray PMS Members of IRAS 09002-4732 \label{acis_mem.tbl}}
\tablehead{
\colhead{Seq} & \colhead{CXOU}  & \colhead{$\log L_x$} & \colhead{$\log N_H$} &  \colhead{2MASS} & \colhead{J} &  \colhead{H} &  \colhead{K} & \colhead{Qual} &   \colhead{IRX} & \colhead{$Age_{JX}$}  \\
 & & \colhead{erg~s$^{-1}$} & \colhead{cm$^{-2}$} & & \colhead{mag} & \colhead{mag} & \colhead{mag} &  & & \colhead{Myr} 
 }
\startdata
      1 & 090105.86-474209.1 & $30.41 \pm 0.37$ & $22.25 \pm 0.30$ & 09010578-4742103 & $16.34 \pm 0.11$ & $14.70 \pm 0.05$ & $13.76 \pm 0.05$ & BAA000 & -0.04 & \nodata \\
      2 & 090106.52-474429.7 & $30.92 \pm 0.21$ & $22.36 \pm 0.14$ & 09010653-4744298 & $16.68 \pm 0.14$ & $15.42 \pm 0.11$ & $14.99 \pm 0.13$ & CAB000 & -0.34 & \nodata \\
      3 & 090107.40-474101.2 & $30.72 \pm 0.32$ & $22.23 \pm 0.37$ & 09010749-4741014 & $17.73 \pm$ \nodata & $15.76 \pm 0.16$ & $14.37 \pm 0.10$ & UCA000 &  0.22 & \nodata \\
      4 & 090109.07-474108.4 & $31.49 \pm 0.70$ & $23.60 \pm 0.34$ & \nodata & \nodata & \nodata & \nodata & \nodata & \nodata & \nodata \\
      5 & 090109.56-474426.0 & $30.29 \pm 0.37$ & $22.23 \pm 0.30$ & 09010955-4744261 & $16.91 \pm 0.20$ & $15.81 \pm 0.14$ & $15.12 \pm 0.16$ & CBC000 &  0.02 & \nodata \\
\enddata
\end{deluxetable}

\begin{deluxetable}{cccccccccc}
\centering
\tabletypesize{\scriptsize} 
\tablehead{
\colhead{Ch1} &  \colhead{Ch2} & \colhead{Ch3} & \colhead{Ch4} &  \colhead{$\alpha_{SED}$} &  \colhead{r} & \colhead{i} &  \colhead{$A_V(JHK)$} &  \colhead{$Age_{ri}$} &  \colhead{Plx} \\
 & & & & & mag & mag & mag & Myr & mas 
}
\startdata
$12.82 \pm 0.05$ & $12.45 \pm 0.10$ & $12.03 \pm 0.14$ & $11.48 \pm 0.10$ & -1.31 & \nodata & \nodata & \nodata & \nodata & \nodata\\
\nodata & \nodata & \nodata & \nodata & \nodata & \nodata & $20.81 \pm 0.09$ & \nodata & \nodata & \nodata \\
$13.18 \pm 0.06$ & $12.77 \pm 0.07$ & $12.33 \pm 0.11$ & $12.07 \pm 0.18$ & -1.42 & \nodata & \nodata & \nodata & \nodata & \nodata\\
$14.32 \pm 0.06$  & $13.40 \pm 0.09$ & $12.56 \pm 0.16$ & $11.85 \pm 0.09$ &  0.03 & \nodata & \nodata & \nodata & \nodata & \nodata\\
$14.44 \pm 0.10$ & $14.08 \pm 0.15$ & \nodata & \nodata & -1.49 & \nodata & $20.96 \pm 0.07$ & \nodata & \nodata & \nodata\\
\enddata
\tablecomments{
This table with 386 lines is available in its entirety in the electronic edition of the Journal.  A few sources are shown here for guidance regarding its form and content. Column definitions: \vspace{3pt} \\
Seq: Sequence number of X-ray selected probable members of IRAS 09002-4732 \vspace{3pt} \\
CXOU:  {\it Chandra} X-ray Observatory Unidentified source  \vspace{3pt} \\ 
$\log L_x$: log X-ray luminosity of sources (assuming distance of 1.7~kpc) obtained with XPHOT \citep{Getman10}. \vspace{3pt} \\
$\log N_H$: log hydrogen column density along the line-of-sight from the median energy  of extracted photons obtained with XPHOT \vspace{3pt} \\
2MASS, $J$, $H$, $K$, $Qual$: 2MASS catalog identifier, $JHK$ magnitudes, and quality flags \vspace{3pt} \\
IRX: Infrared excess derived from 2MASS color-color and color magnitude  diagrams \vspace{3pt} \\
$Age_{JX}$: Stellar age estimated from 2MASS and {\it Chandra} magnitudes as described by \citet{Getman17} \vspace{3pt} \\
Ch1, Ch2, Ch3, Ch4: Spitzer IRAC magnitudes in Channels 1 (3.5$\mu$m), 2 (4.5$\mu$m), 3 (5.6$\mu$m), and 4 (8.0$\mu$m) bands \vspace{3pt} \\
$\alpha_{SED}$: Apparent infrared spectral energy distribution slope derived from IRAC photometry \vspace{3pt} \\
$i$, $i$: VPHAS+ $r$ and $i$ band magnitudes \vspace{3pt} \\
$A_V(JHK)$: Visual absorption estimated from 2MASS $JHK$ colors \vspace{3pt} \\
$Age_{ri}$: Stellar age estimated from VPHAS+ $r-i$ color-magnitude diagram \vspace{3pt} \\
Plx: Gaia parallax 
}
\end{deluxetable}

\newpage

\begin{deluxetable}{cccccccrrrrrcc}
\centering
\tabletypesize{\tiny} 
\tablecaption{{\bf MIRES sources from Povich et al. (2013) undetected in X-rays}  \label{MIRES.tbl}}
\tablehead{
\colhead{Seq} & \colhead{IRAC} & \colhead{RA} & \colhead{Dec} & \colhead{J} & \colhead{H} & \colhead{K} & \colhead{Ch1} & \colhead{Ch 2} & \colhead{Ch 3} & \colhead{Ch 4} & \colhead{$\alpha_{SED}$} & \colhead{$\alpha_N$} & \colhead{Stage} }
\startdata
387 &  G268.2943-00.9270 & 135.26 & -47.69 &  \nodata              & 15.00$\pm$0.07 & 13.43$\pm$0.05 & 11.88$\pm$0.05 & 11.16$\pm$0.05 & 10.74$\pm$0.10 & 10.11$\pm$0.10 & -0.77$\pm$0.12 &   4 &  -1 \\
388 &  G268.2940-00.9156 & 135.28 & -47.68 &  \nodata              & \nodata               &  \nodata              & 13.85$\pm$0.06 & 13.22$\pm$0.09 & 12.56$\pm$0.15 & 11.45$\pm$0.17 & -0.19$\pm$0.18 &   4 &  -1 \\ 
399 &  G268.3193-00.9338 & 135.28 & -47.71 & 16.62$\pm$0.14 & 15.04$\pm$0.06 & 14.32$\pm$0.07 & 13.19$\pm$0.07 & 12.74$\pm$0.09 & 12.13$\pm$0.16 & 11.27$\pm$0.18 & -0.71$\pm$0.20 &   4 &  -1 \\ 
400 &  G268.2619-00.8819 & 135.28 & -47.64 & 14.13$\pm$0.05 & 12.73$\pm$0.05 & 11.61$\pm$0.05 & 10.22$\pm$0.05 &   9.83$\pm$0.05 &   9.28$\pm$0.10 &   8.47$\pm$0.10 & -0.89$\pm$0.12 &   4 &  ~2 \\ 
401 & G268.3562-00.9516 & 135.30 & -47.75 &  \nodata              & 15.77$\pm$0.14 & 14.88$\pm$0.12 & 13.45$\pm$0.07 & 12.04$\pm$0.05 & 10.73$\pm$0.10 &   9.96$\pm$0.10 &  1.30$\pm$0.14 &   4 &  ~1 \\ 
\enddata
\tablecomments{
This table with 55 lines is available in its entirety in the electronic edition of the Journal.  A few sources are shown here for guidance regarding its form and content.
Information in this table is from \citet{Povich13} except for $\alpha_{SED}$, the infrared spectral slope calculated from $\alpha_N$ photometric IRAC points.  
Stage = 1 corresponds to Class 0/I protostar; Stage = 2 corresponds to Class II/III PMS star; Stage = -1 corresponds to an ambiguous status.  These sources are not detected in the {\it Chandra} field. }

\end{deluxetable}

\clearpage \vfill
 
\section{Pre-Main Sequence Members of IRAS 09002-4732} \label{members.sec}

\subsection{X-ray Sources and Stellar Counterparts} \label{catalog.sec}

Table~\ref{candx.tbl} lists the 1034 candidate X-ray sources in the IRAS 09002-4732 region obtained using the procedures outlined in \S\ref{datameth.sec}.   This candidate X-ray source list is purposefully very sensitive and contains spurious noise features.  They are kept here because, in our experience, some of the weaker candidates have clear infrared candidates and are likely PMS members of the region.  Spurious noise features, and real X-ray sources associated with background extragalactic objects, will rarely have optical and infrared (OIR) counterparts.  

Omitting the handful of X-ray candidates with $<2.0$ net counts, this X-ray catalog was matched to four OIR catalogs of point sources\footnote{
Three catalogs were accessed from the online Vizier catalog service provided by the University of Strasbourg Centre des Donn\'ees Stellaires (CDS) as follows: VST Photometric H$\alpha$ Survey of the Southern Milky Way and Bulge (VPHAS+) at \url{http://cdsarc.u-strasbg.fr/viz-bin/Cat?II/341}, Gaia Data Release 2 (DR2) at \url{https://vizier.u-strasbg.fr/viz-bin/VizieR-3?-source=I/345/gaia2}, Two Micron All-Sky Survey (2MASS) at \url{http://vizier.u-strasbg.fr/cgi-bin/VizieR?-source=B/2mass}.  The Spitzer IRAC source catalog was obtained from the electronic edition of the MYStIX Infrared-Excess Source (MIRES) by \citet{Povich13}.
}:  
VPHAS+-DR2, Gaia-DR2 in a broad optical band, 2MASS in the $JHK$ near-infrared bands, and Spitzer IRAC in four mid-infrared bands in the range $3.6-8.0$ $\mu$m. In addition we checked the 264 JHK stars listed by Apai et al. (2005) in the inner region of IRAS 09002-4732.  

The search radius for finding counterparts from X-ray centroids (Table~\ref{candx.tbl}) was chosen to be 1\arcsec\/ within the central part of the ACIS field, and 2\arcsec\/ for off-axis angles exceeding 4\arcmin. Among 1034 X-ray candidate sources, 166, 110, 244, and 201 have matches with sources from the VPHAS+, Gaia, 2MASS, and IRAC catalogs, respectively.  Taking duplicates into account, 300 of the X-ray candidate sources have OIR counterparts.  We have confirmed that the celestial positions of the X-ray sources and 2MASS sources are well-aligned with no measurable ($<0.1$\arcsec) systematic offset.  

An additional 86 X-ray sources without OIR counterparts are spatially clustered in the central part of the region. The combined sample of the 300 and 86 X-ray sources is given in Table~\ref{acis_mem.tbl}. Table~\ref{MIRES.tbl} gives a further subsample of 55 non-X-ray disk-bearing stars from Povich et al. located within the {\it Chandra} field.  These two subsamples are described later in this section.  

The total sample of 441 ($300+86+55$) young stellar candidates are then subject to three stages of statistical analysis.
First, their spatial distribution is segmented into the sum of ellipsoidal isothermal spheres, plus a uniform background population following the methodology of  \citet{Kuhn14} and \citet{Kuhn18}.  This is performed with a maximum likelihood mixture model where the optimal number of ellipsoids is obtained using the penalized likelihood Akaike Information Criterion.  Second, the total population of each cluster is estimated from scaling the truncated X-ray Luminosity Function (XLF) to the more complete XLF of the ONC, as described by \citet{Kuhn15}.  Third, the age of each stellar cluster is estimated in two fashions:  $Age_{JX}$ chronometer for PMS stars that uses absorption-corrected X-ray luminosities and $J$ magnitudes \citep{Getman14}; and the traditional $Age_{ri}$ chronometer based on the optical color-magnitude diagram.

Cautions on the completeness and reliability of this catalog of X-ray selected sources are needed:
\begin{enumerate}
\item All OIR catalogs in the central region of IRAS 09002-4732 are subject to spatially variable nebulosity from the H~II region and dense dusty molecular filaments \citep{Apai05}.  This gives patchy catalog coverage even from uniform OIR surveys such as VPHAS+, 2MASS and UKIDSS.  These problems are evidenced in the low quality detections or nebular contamination reported in 2MASS quality flags.  More complete OIR coverage would give additional counterparts to the candidate X-ray sources\footnote{
An example of this effect is that six additional candidate X-ray sources have counterparts within 1\arcsec\/ in the list of 268 $JHK$ sources from high-resolution VLT/ISAAC images presented by \citet{Apai05} available from the Vizier database at \url{http://vizier.u-strasbg.fr/viz-bin/VizieR?-source=J/A+A/434/987}.  These sources are: (ISAAC Seq \#, CXOU) = (306, 090145.23-474513.6), (533, 090156.03-474512.2), (544, 090156.69-474535.7), (560, 090157.32-474542.4), (632, 090201.43-474537.5), (652, 090202.45-474525.1).  These stars have $K \sim 13-15$ mag and are absent from the 2MASS catalog due to the confusing nebulosity and absorption of the region discussed by Apai et al.  We do not include these sources in our analysis as this infrared catalog is incomplete in flux and spatial coverage.  \label{apai.footnote}
}.
\item In the X-ray band, the central $10\arcsec-15$\arcsec\/ of the cluster suffers reduced exposure due to the 11\arcsec\/ physical gaps between the CCDs of the ACIS-I array.  These are shown in the blue outlines of the chips in Figure~\ref{Herschel-Chandra.fig}.   This is partially overcome by the 16\arcsec\/ Lissajous-shaped dithering pattern of the satellite aspect system \citep{Chandra18}.  For the sum of the three observations in Table~\ref{CXO.tbl}, the resulting effective exposure in the inner few arcseconds of the cluster is roughly half that of the rest of the region, resulting in a localized deficit of sources in the innermost cluster core.  

\item Random Galactic field stars or extragalactic sources can lie within $1\arcsec-2$\arcsec\/ radii of X-ray sources.  Based on simulations of contaminant populations for the MYStIX $Chandra$ observation of the nearby RCW~38 cluster, $\sim 200$ of the 1034 candidate X-ray sources (Table~\ref{candx.tbl}) are likely true X-ray sources unrelated to the IRAS 09002-4732 star forming region \citep{Broos13}.  However, the majority are quasars extremely faint in the OIR bands (particularly after extinction by cloud material) and will not appear in our list of likely pre-main sequence members. Following the SFiNCs analysis of \citet{Getman17} involving shifts of the ACIS field in different directions, we estimate that $\sim 8$\% of the OIR counterparts are stellar contaminants, consistent with the simulations of Broos et al.
   
\end{enumerate}

Fifty-five infrared sources undetected in X-rays that lie within the {\it Chandra} IRAS 09002-4732 field, typically having seven bands of photometry from 2MASS and Spitzer IRAC, were identified in the MIRES catalog of \citet{Povich13} as sources with spectral energy distributions indicative of disk-bearing PMS stars; these are listed in Table~\ref{MIRES.tbl}. The criteria selecting these objects from the vast majority of older stars in the infrared catalogs is stricter than usually applied in studies of nearby young stellar object populations to reduce contamination by the many dusty red giants that contaminate deep surveys in the Galactic Plane.  Twenty-one of these sources have infrared spectral energy distributions characteristic of protostars, denoted Stage 1 by Povich et al. The MIRES protostars in the IRAS 09002-4732 vicinity have a distribution of $2-8$~$\mu$m luminosities similar to the typical protostars in the Orion cloud; but this distribution is missing a high-luminosity tail seen in Orion \citep{Megeath12}.

\begin{figure}[h]
\centering
\includegraphics[width=0.75\textwidth]{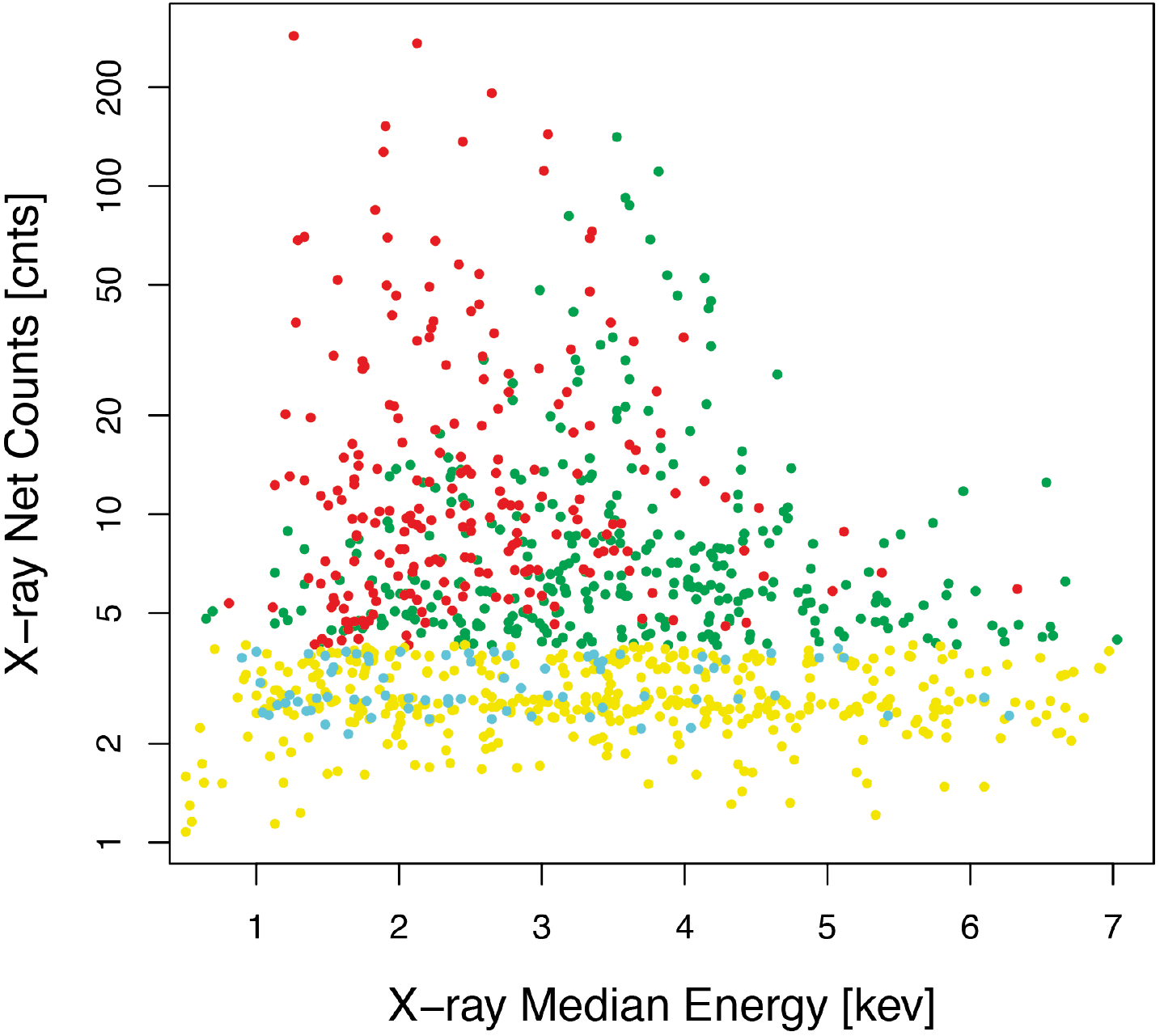}
\caption{Sample of 1034 candidate X-ray sources in the IRAS 09002-4732 {\it Chandra} field with three properties: net counts $C_{net}$ in the $0.5-8$~keV band; median energy $ME$ (in keV) of the net counts; and existence of OIR counterparts.  Candidate sources with counterparts are plotted in red if $C_{net}>4$ and in cyan if $2<C_{net}<4$.  Candidate sources without counterparts are plotted in green  if $C_{net}>4$ and in yellow if $C_{net}<4$.  
\label{xsrc_me.fig}}
\end{figure}

Figure~\ref{xsrc_me.fig} is an `X-ray color-magnitude diagram' that provides information about the reliability of the 1034 X-ray candidate sources.  The red and green symbols show sources with $\geq 4$ net counts in the ACIS image which, in most cases, is a statistically significant X-ray source.  From experience with X-ray color-magnitude diagrams from the MYStIX and SFiNCs surveys \citep{Kuhn13, Broos13, Getman17}, the sources shown in red with OIR counterparts are expected to be PMS members of the region. The green hard-spectrum sources without OIR counterparts are composed of both contaminant populations (extragalactic and Galactic) and PMS stars;  in the latter case, the stars are missing from the OIR catalogs due to H~II region nebulosity and/or heavy cloud extinction. (Note that PMS stars with X-ray median energies ranging from 2.5 to 5~keV will have $K$ band extinctions ranging from $\sim 1$ to $\sim 10$ mag;  Getman et al. 2010.)    Among the sources with $2-4$ net counts, the majority of cyan ones are likely PMS stars while the yellow ones are either spurious noise fluctuations from the X-ray image or contaminant populations.  From this analysis, we conclude that the X-ray membership sample is likely to include several spurious sources (cyan symbols) and miss some real members (green symbols).  These deficiencies can be alleviated if deeper X-ray and high-resolution infrared images of the $17\arcmin \times 17$\arcmin\/ region are obtained.  

\begin{figure}[h]
\centering
\includegraphics[width=0.85\textwidth]{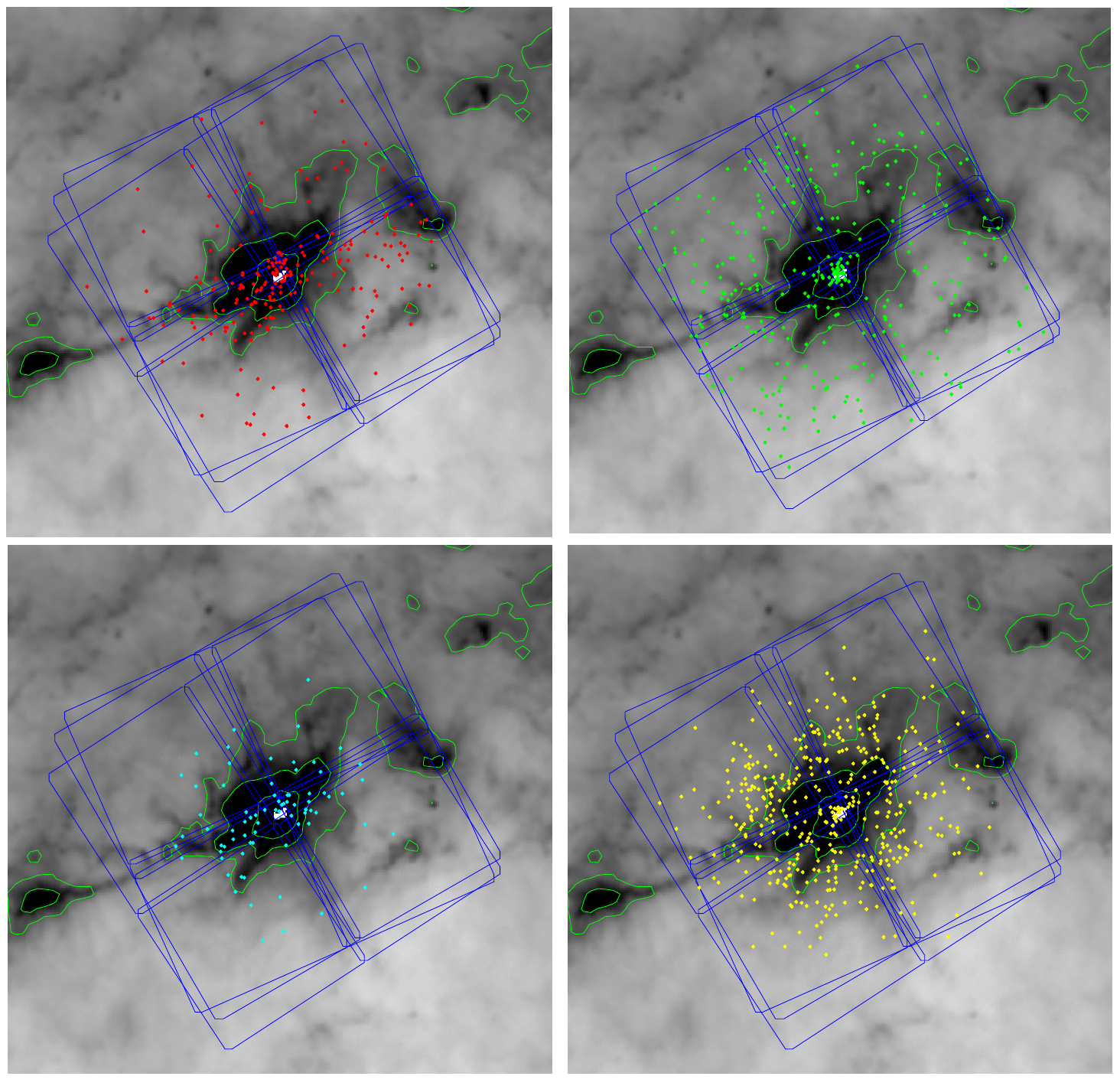}
\caption{Spatial distribution of four subsamples of candidate X-ray sources superposed on a Herschel SPIRE 250~$\mu$m map of the IRAS~09002-4732 region.   The green contours represent infrared surface brightness levels of 500, 1000, and 3000 MJy/sr; these show the shape of the dusty cloud structures.  The candidate X-ray sources are divided into subsamples and colors following Figure~\ref{xsrc_me.fig}.  See the text for details.  \label{CXO_cts.fig}}
\end{figure} 

The spatial distribution of all X-ray sources detected in the IRAS~09002-4732 field is shown in Figure~\ref{CXO_cts.fig} superposed on a 250~$\mu$m map of the cool dust from the Herschel-SPIRE instrument.    The four panels show subsamples using the same color scheme as in Figure~\ref{xsrc_me.fig}.  The left panels show X-ray candidate sources with OIR  counterparts, while the right hand panels show sources without counterparts.  The upper left panel shows the reliable subsample of stronger X-ray sources ($>4$ net counts) with OIR counterparts include sources both concentrated toward the center of the rich cluster and distributed widely around the molecular cloud.  The lower left panel showing tentative X-ray sources with OIR counterparts (cyan) have a similar spatial distribution, suggesting they are mostly real members.  

The candidate X-ray sources without OIR counterparts are shown in the right panels of Figure~\ref{CXO_cts.fig}; the top panel has sources with $>4$ net counts and the lower panel has sources with $2-4$ net counts.  Two distinctive spatial groups are present: several dozen X-ray sources are densely concentrated in the cluster core, while most are scattered uniformly across the ACIS field.   A random distribution is expected from background quasars or spurious noise sources.  The concentrated group most likely reveals true members of the PMS cluster despite the absence of OIR counterparts or high statistical significance in the X-ray image.  We find a concentration of 86 X-ray sources without OIR counterparts  projected within a chosen radius 1.5\arcmin\/ of a central location at $(\alpha, \delta) = (135.47435^\circ, -47.72820^\circ)$. This corresponds to a cluster with diameter 1.5~pc;  this encompasses the molecular cores, infrared reflection nebulosity, and many of the infrared-excess stars identified by \citet{Apai05}. 

\vfill 

\subsection{The IRAS 09002-4732 Member Sample} \label{PMS_sample.sec}

For the present study, we construct a sample of 441 probable stellar `members' of the IRAS 09002-4732 region as the union of three subsamples:
\begin{enumerate}
\item 300 candidate X-ray sources with $>2$ X-ray net counts that have a counterpart within 1\arcsec-2\arcsec\/  in any of the four OIR catalogs, listed in Table~\ref{acis_mem.tbl};
\item 86 candidate X-ray sources with $>2$ X-ray net counts that lie projected against the central part of the cluster (a circle with diameter 1.5 pc) and lack an OIR counterpart, also listed in Table~\ref{acis_mem.tbl};
\item 55 members of the MIRES catalog with infrared excesses located within the {\it Chandra} field that lack X-ray counterparts, listed in Table~\ref{MIRES.tbl}.
\end{enumerate}
The two subsamples based on candidate X-ray sources are numbered sequentially $1-386$ by increasing right ascension in Table~\ref{acis_mem.tbl}.  The infrared selected subsample is numbered sequentially $387-441$ in Table~\ref{MIRES.tbl}.    

The spatial distribution of IRAS 09002-4732 members is placed into a wider context in Figure~\ref{Herschel-Chandra.fig} that includes both the IRAS~09002-4732 region and the MYStIX target H~II region RCW 38 to the west.  MIRES disk-bearing candidates from \citet{Povich13} are shown as cyan (Class~II/III) and red (protostars) symbols.  Recall that both RCW~38 and IRAS~09002-4732 have bright infrared nebulosity from H~II region emission lines and heated dust;  this considerably reduces the sensitivity of the infrared surveys causing the MIRES infrared-excess catalog to be very incomplete.  

\begin{figure}[ht]
\centering
\includegraphics[width=0.75\textwidth]{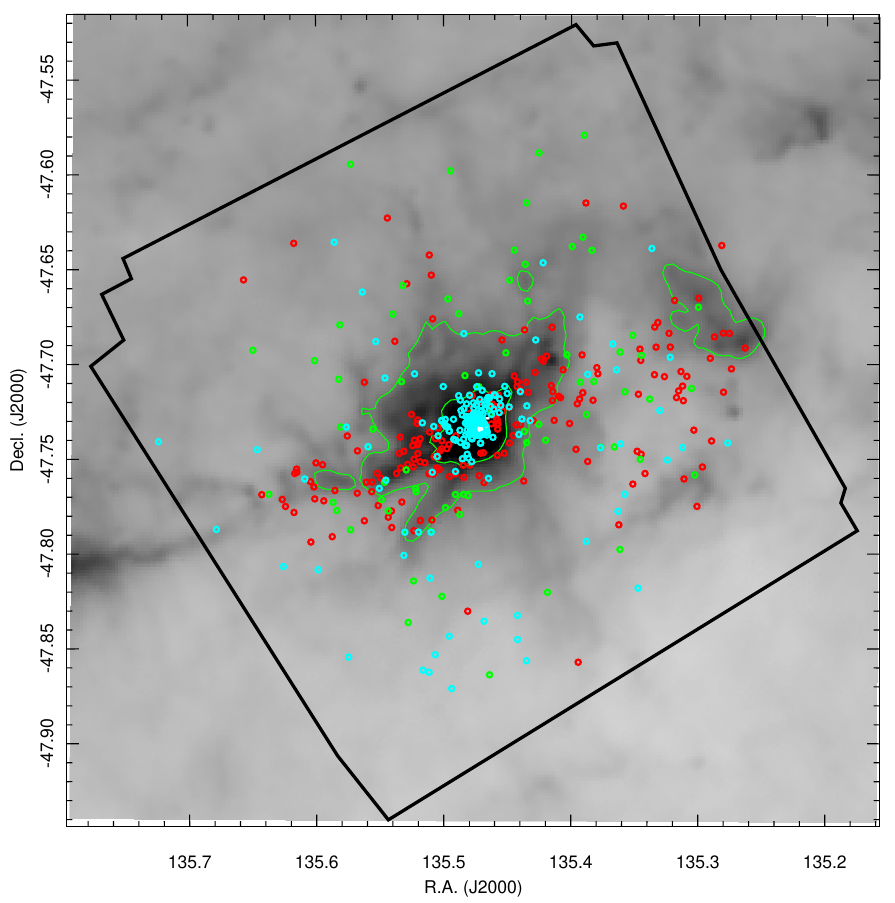}
\caption{Spatial distribution of 386 X-ray selected PMS members coded by disk infrared emission as an indicator of stellar age:  disk present (red symbols); disk absent (green); and undetermined (cyan).  \label{acis_disk.fig}}
\end{figure}

\vfill 
\subsection{Star formation along the molecular filament} \label{filament.sec}

Examination of Figure~\ref{Herschel-Chandra.fig} immediately reveals that the spatial distribution of the MIRES stars, selected to have infrared spectral energy distributions consistent with low-mass protostars, is strongly elongated along a northwest-southeast axis (approximately along P.A. 110$^\circ$ East of North).  The northwest direction points toward the large RCW~38 cloud, and the southeast direction is along a thin dusty filament with a small cloud condensation a few arcminutes outside the {\it Chandra} field of view.  This strongly suggests that current star formation is concentrated along a filament several parsecs long.  On a smaller scale of  $0.5-1$~parsec around the IRAS 09002-4732 cluster core, the VLT-ISAAC $JHK$ image presented by \citet{Apai05} shows dense obscuring cloud structure along the same northwest-southeast axis.   

A similar spatial trend is present in the ages of X-ray selected members, as inferred from the presence or absence of an infrared-emitting inner circumstellar disk.  Recall that X-ray selection of PMS stars is largely independent of disks, as it arises from magnetic reconnection flares.  Figure~\ref{acis_disk.fig} shows the 386 ACIS sources with colors representing the presence of disk emission in the spectral energy distributions obtained from 2MASS $JHK$ and Spitzer IRAC $3.6-8.0$~$\mu$m photometry.  A disk is inferred to be present when the infrared power law slope exceeds $-1.9$ \citep{Richert18}.  The disk-bearing stars show a clear tendency to lie along the northwest-southeast axis, while the disk-free or undetermined cases (due to inadequate infrared photometry in the nebular region) exhibit a more spherical distribution.  

The MIRES stars in the filament are not strongly concentrated into groups or clusters, but rather appear to be reflect a condition of distributed star formation.  This differs from other situations in the massive MYStIX star forming regions where embedded small clusters are common in the vicinity of a revealed rich cluster \citep{Kuhn14, Getman18}.  There are other cases where infrared-excess stars are mostly found in groups, such as the large InfraRed Dark Cloud G53.2 at a distance similar to IRAS 09002-4732 \citep{Kim15}.  

\subsection{Identifying clusters} \label{cluster.sec}

\begin{figure}[h]
\centering
\includegraphics[width=0.8\textwidth]{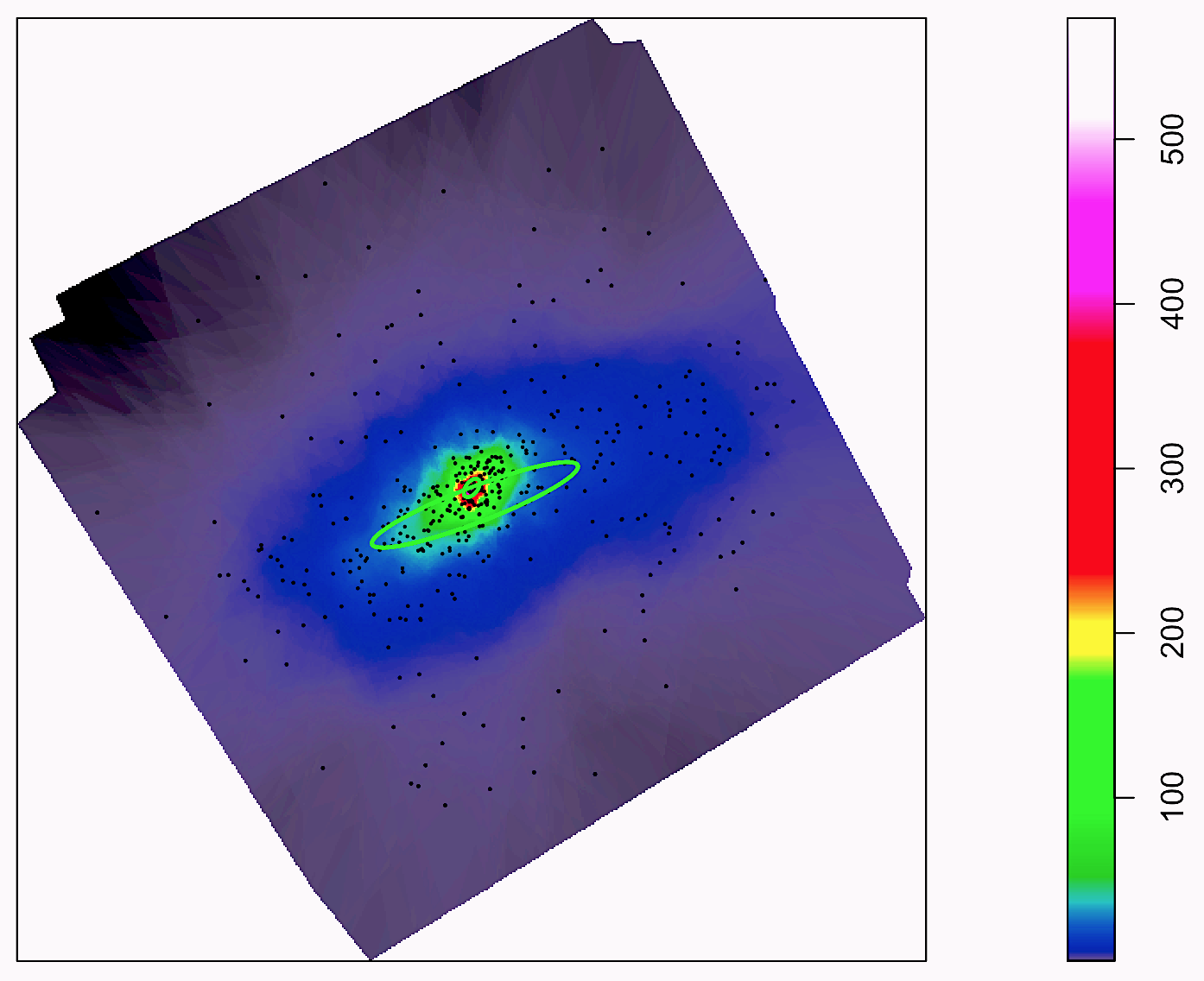}
\caption{Maximum likelihood mixture model fit decomposing the spatial distribution of 441 PMS members into two components. The color map shows an adaptively smoothed image  with a color bar in units of observed stars pc$^{-2}$.  Black dots indicate the individual stars, and the green ellipses show the core radii of the two ellipsoidal components.  \label{Two_ellipse.fig}}
\end{figure}

As mentioned in \S\ref{catalog.sec}, we apply a statistical procedure developed by \citet{Kuhn14} for identifying stellar clusters in two-dimensional spatial point distributions.   The method is based on maximum likelihood estimation of a mixture model where, instead of the usual Gaussian shape, we assume each cluster can be represented by an isothermal ellipsoid with a central surface density $\sigma_0$, core radius $r_c$, ellipticity $\epsilon$, and orientation $\theta$.   The ellipticity is defined to be $(a-b)/a$ where $a$ and $b$ are the major and minor axes respectively.  The optimal number of clusters $k$ of the model is obtained by maximizing the Akaike Information Criterion, a widely used penalized likelihood measure for selecting parsimonious models.   Ellipsoidal components can be disjoint, overlapping, or with a small component encompassed within a larger one.  

The best fit model for the 441 spatial distribution of X-ray and infrared selected members of the IRAS 09002-4732 region (\S\ref{PMS_sample.sec}) is shown in Figure~\ref{Two_ellipse.fig}.   The majority stars are members of a compact cluster centered at (09$^h$01$^m$54.2$^s$, -47$^\circ$43$^\prime$48$^{\prime\prime}$).  This is 23\arcsec\/ north of the dominant O7 star and the radio continuum peak, lying between dense filaments S2 and S3 that obscuring the cluster center  \citep[][ and Appendix \ref{core.sec} below]{Apai05}.  This cluster has major and minor axis core radii $0.08 \times 0.14$~pc ($\epsilon = 0.43$) with $\theta = 143^\circ$ east of north.  The remaining stars are members of a more elongated, and lower density cluster centered at (09$^h$01$^m$53.7$^s$, -47$^\circ$44$^\prime$16$^{\prime\prime}$).  This cluster has major and minor axis core radii $1.45 \times 0.23$ pc ($\epsilon = 0.84$) with $\theta =  111^\circ$. Judging from the results of the SFiNCs Monte Carlo simulations \citep[][their Figure~5]{Getman18b}, the formal fractional errors on the inferred cluster core radii and ellipticities in IRAS~09002-4732 should not exceed 20\%. The elongated cluster component clearly delineates the findings in \S\ref{filament.sec} that star formation is concentrated in a northwest-southeast molecular filament \S\ref{filament.sec}. There is no statistically significant substructure in either grouping, and there is no significant uniformly distributed unclustered stellar population from an earlier generation of star formation. 

The  central surface density of the compact cluster is $\sim 20$ times higher than the elongated cluster but its spatial extent is smaller.  It is thus difficult to decide on their relative populations, or to assign specific stars to one or the other component.   We roughly estimate that 80\% of the stars belong to the compact cluster and 20\% belong to the protostellar population.  

\section{Properties of clusters and individual stars} \label{properties.sec}

\subsection{Distance} \label{dist.sec}

Historical estimates of the distances to Galactic Plane star forming regions typically vary considerably.  For IRAS 09002-4732, \citet{Lenzen91} adopts 1.8 kpc, \citet{Ghosh00} have a best-fit of 1.4~kpc to the embedded O7 star, and \citet{Apai05} assume 1.3 kpc.  Our MYStIX project assumed 1.7 kpc for RCW 38 \citep{Feigelson13} following the review by \citet{Wolk08}.  But today, for a small fraction of member stars that are sufficiently bright in the optical band, the DR2 catalog of ESA's Gaia satellite provides direct measurements of parallactic distance \citep{GaiaCollab16, GaiaCollab18}. 

\begin{figure}[h]
\centering
\includegraphics[width=0.9\textwidth]{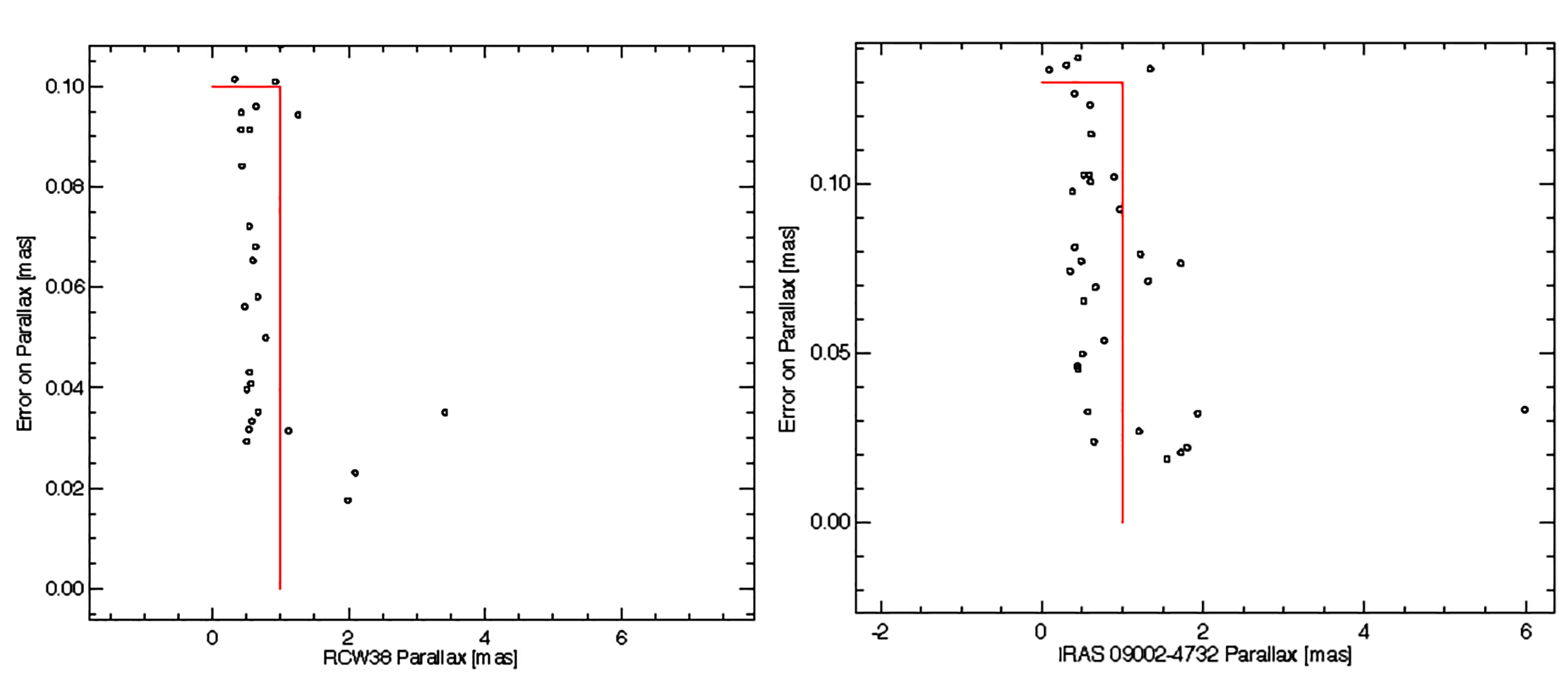}
\caption{Parallax measurements of PMS members of RCW~38 (left) and IRAS~09002-4732 (right) from the Gaia DR2 catalog.  To guide the eye, the vertical red lines show the parallax value of 1~mas  and the horizontal red lines show errors of 0.10 mas (left) and 0.13 mas (right), respectively.  These values are not corrected for any systematic shift.   \label{Gaia.fig}}
\end{figure}

Gaia parallaxes to PMS members of IRAS 09002-4732 and its neighboring RCW~38 cluster are shown  Figure~\ref{Gaia.fig}, restricted to stars with parallax errors less than $\sim$0.1~mas and $\sim$0.13~mas for the two regions respectively.   The sample includes 18 members of RCW~38 from the MYStIX Probable Complex Members \citep{Broos13} and 20 members of IRAS~009002-4732 identified here.  The parallax distribution appears to have outliers, perhaps due to contamination by nebulosity.  A concentration is seen around $0.55-0.60$~mas corresponding to $1.7-1.8$~kpc.  For consistency with our MYStIX study, we adopt a distance of 1.7 kpc for this study.  

\subsection{Extinction} \label{extinct.sec}

The IRAS 09002-4732 cluster is known to be highly extincted: the stars and emission nebulosity is obscured by dense dusty filaments and globules with characteristic obscuration around $A_V \simeq 20$~mag, individual PMS stars exhibit extinction ranging from $A_V \simeq 3$ to 30~mag, and a 9.7$\mu$m absorption feature in the ultracompact H~II region corresponds to $A_V \simeq 20$~mag \citep{Apai05}.  \citet{Lenzen91} infer a higher absorption of $A_V > 45$~mag, and possibly $A_V > 70$~mag, from the spectral energy distribution of the dominant O7 star.  

\begin{figure}[h]
\centering
\includegraphics[width=0.45\textwidth]{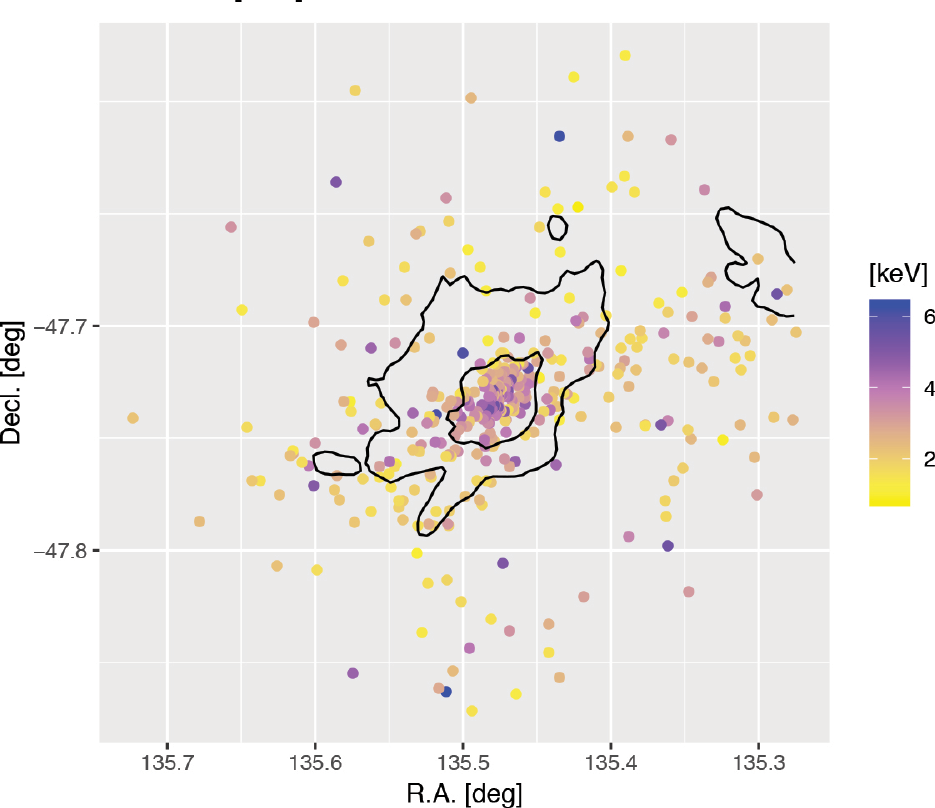}
\includegraphics[width=0.48\textwidth]{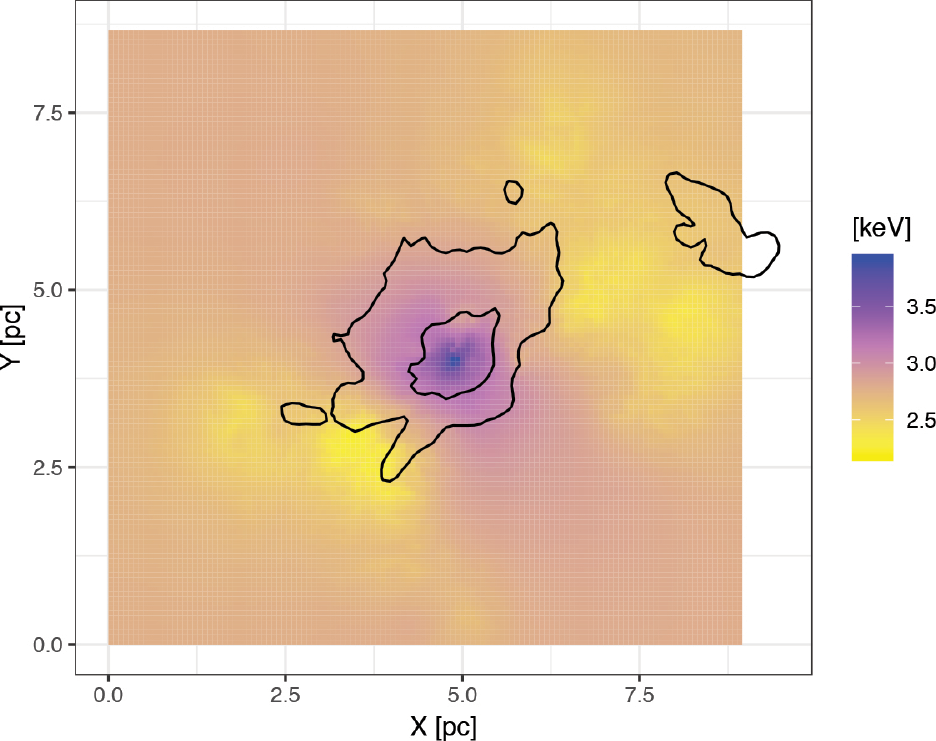}
\caption{Spatial distribution of interstellar extinction to X-ray selected members of IRAS~09002-4732 based on observed Median Energy of photons from each source.   The left panel shows individual source median energies with celestial coordinates, while the right panel shows an adaptively smooth mean value with parsec coordinates assuming a distance of 1.7~kpc.  The contours are based on surface brightness of the Herschel 250~$\mu$m map.    \label{CXO_abs.fig}}
\end{figure}

X-ray emission in the {\it Chandra} $0.5-8$~keV band is sensitive to line-of-sight absorption by interstellar material.  Here the effect is mostly due to bound-free photoionization of gas-phase metal atoms, not to scattering and absorption effects of solid grains as in the OIR bands \citep{Wilms00}.  \citet{Getman10} have calculated a conversion between this absorption of softer X-rays and the observed median energy of a PMS star, assuming a typical PMS flare spectrum and the spectral response of the {\it Chandra} mirror and ACIS detector, giving estimates of hydrogen column densities $\log N_H$ for individual X-ray sources (Table~\ref{acis_mem.tbl}).  Most stars lie in the range $21.8 < \log N_H < 23.3$ cm$^{-2}$ with median value of 22.4 cm$^{-2}$.   Adopting a gas-to-dust ratio $N_H / A_V = 2 \times 10^{21}$ cm$^{-2}$~mag$^{-1}$ \citep{Zhu17}, this corresponds to a range of $3 \lesssim A_V \lesssim 100$~mag with median value $A_V \simeq 13$~mag. 

The absorption to the X-ray selected IRAS 09002-4732 members exhibits a strong spatial gradient shown in Figure~\ref{CXO_abs.fig}.  The right panel shows  an adaptively smoothed  map of the mean Median Energy values\footnote{This map is calculated with the function {\it adaptive.density} in CRAN package {\it spatstat} within the public domain R statistical software environment \citep{Baddeley15}.}.  Stars on the periphery of the {\it Chandra} field, with projected distances exceeding $\sim 1$~pc show typical absorptions around $A_V \simeq 10-20$ mag, considerably lower than stars concentrated in the cluster center that reach $A_V \simeq 50$ mag.   

\begin{figure}[b]
\centering
\includegraphics[width=0.9\textwidth]{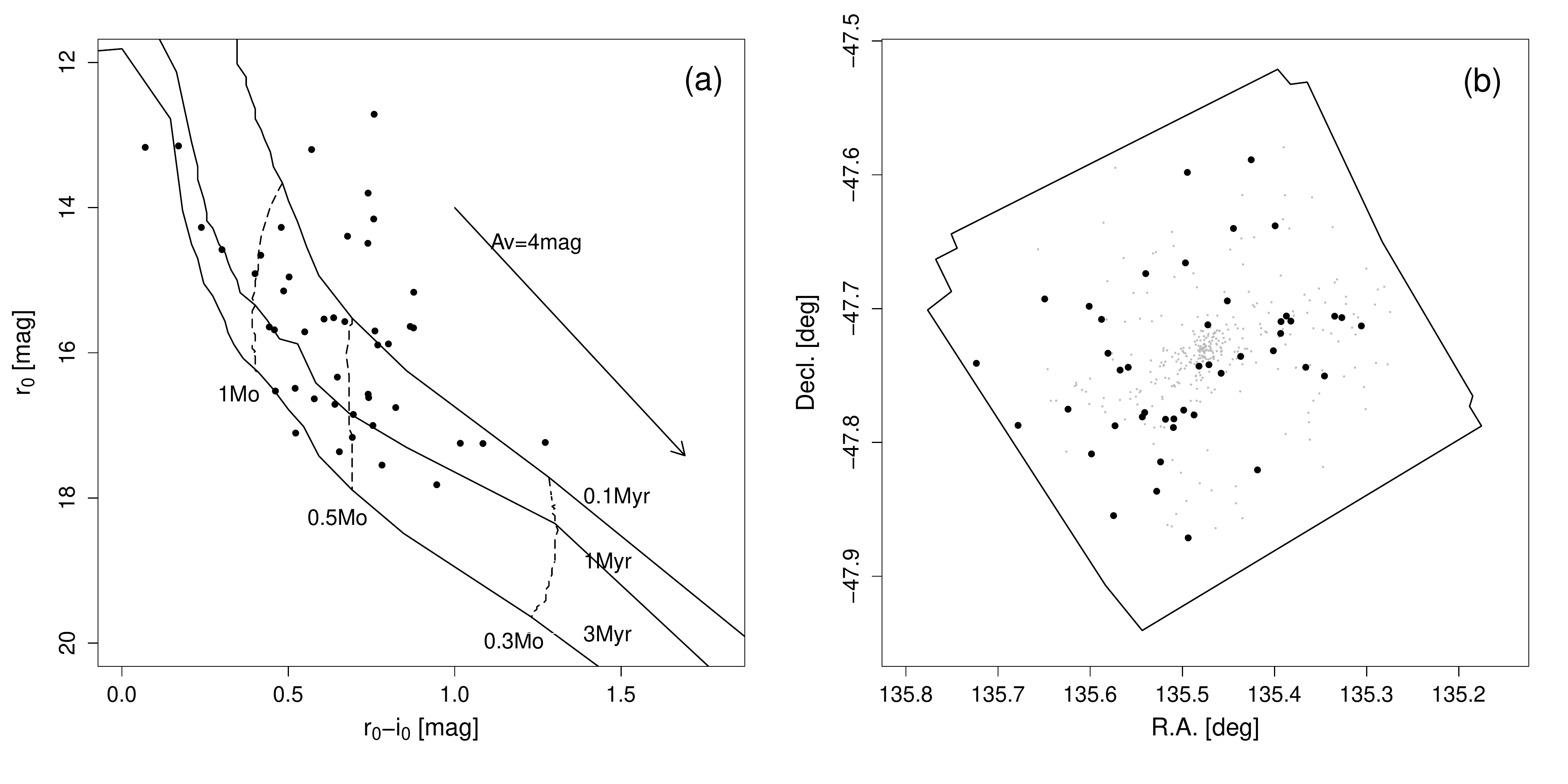}
\caption{Optical color-magnitude diagram (a) and spatial distribution (b) for 45 stars around the IRAS 09002-4732 cluster for estimation of stellar ages.  \label{ri_cmd.fig}}
\end{figure}

\subsection{Age} \label{age.sec}

We provide age estimates for individual stars using two different chronometers for subsamples of the IRAS 09002-4732 stellar population:
\begin{enumerate}
\item \citet{Getman14} presents the $Age_{JX}$ estimator for individual PMS stars based on absorption-corrected X-ray luminosity,  dereddened $J$ magnitudes, distance, and standard PMS evolutionary tracks \citep{Siess00}.  As this requires both high-quality X-ray and near-infrared photometry, it can be calculated for only 22 stars located around the central cluster.  For this subsample, the median age is $Age_{JX} = 0.8$~Myr.   

\item Forty-five  stars with low absorption that are dispersed around the central cluster have optical band photometry from the VPHAS+ survey.  Their $r_0$ $vs.$ $r_0-i_0$ color-magnitude diagram is shown in Figure~\ref{ri_cmd.fig} along with the \citet{Siess00} tracks and their spatial distribution.  Here we see a range of ages from $<0.1$~Myr to 3~Myr with median around 0.7~Myr.   Some of these are widely distributed and others probably lie in the molecular filament; stars in the cluster core are underrepresented due to difficulties of high obscuration and bright nebulosity.  
\end{enumerate}

We thus estimate an average age of 0.8~Myr for the central cluster of IRAS 09002-4732 with a range of $<0.1$ to $\sim 3$~Myr for individual stars.  The protostars distributed along the molecular filament are undoubtedly younger, and it is reasonable to assign them an age around 0.1~Myr.   

Many young clusters exhibit an age gradient in $Age_{JX}$ in the sense that the cluster cores appear younger (that is, formed stars later) than the peripheries of the clusters \citep{Getman18}.  There is some evidence for this from Figure~\ref{acis_disk.fig} where the older disk-free stars (green symbols) are more widely distributed than the concentrated cluster.  However, due to nebulosity and obscuration, insufficient infrared photometry is available in the central region to check whether the disk fraction is higher in the cluster core.  

\begin{figure}[b]
\centering
\includegraphics[width=0.55\textwidth]{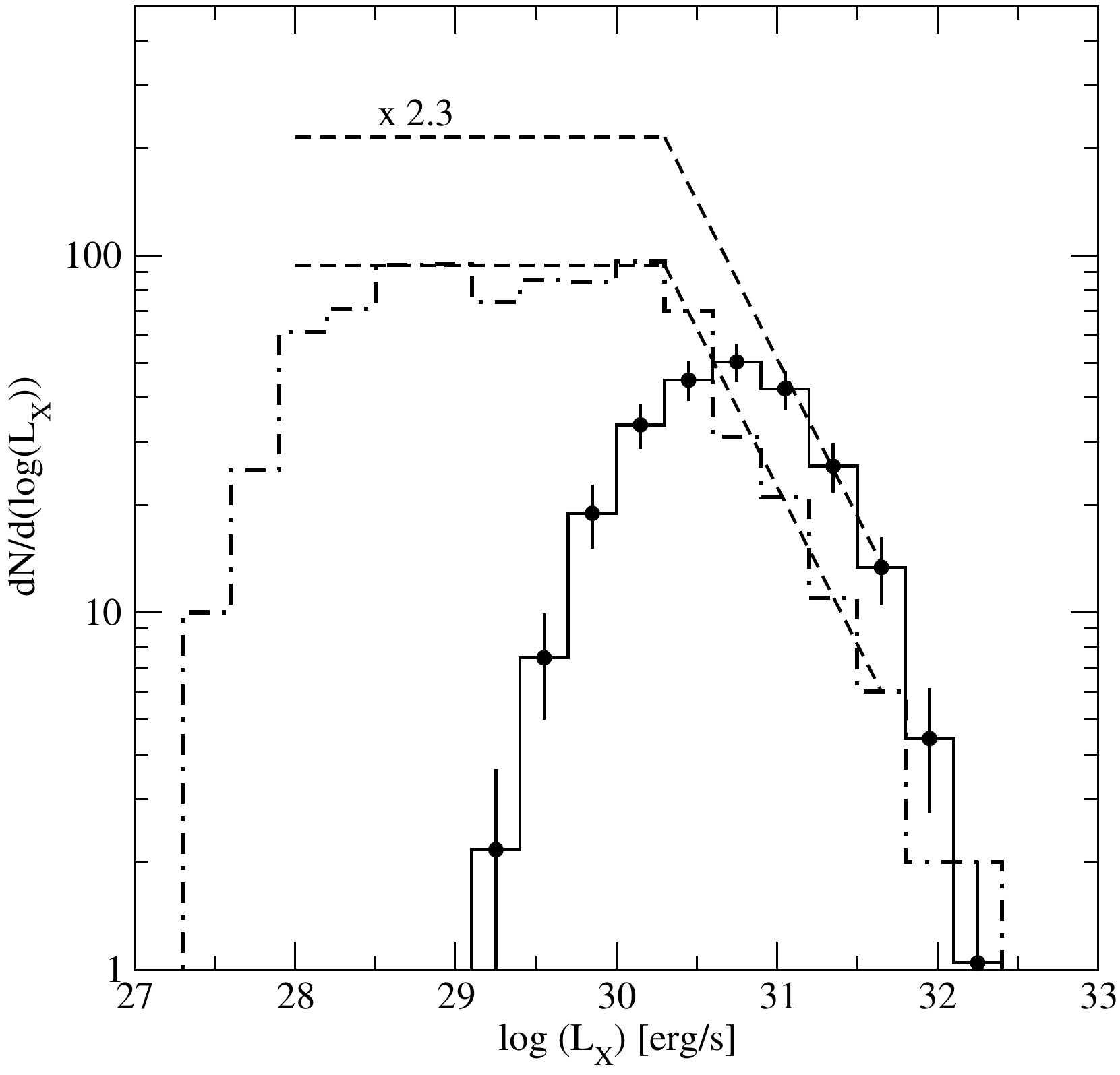}
\caption{Calibration of the IRAS 09002-4732 X-ray luminosity function (solid histogram) to the XLF of the Orion Nebula Cluster (dot-dashed histogram). The two dashed lines show the offset in population between the two clusters.   \label{xlf.fig}}
\end{figure}

\subsection{Total stellar population} \label{pop.sec}

The X-ray luminosity functions (XLFs) of young ($t \leq 5$~Myr) PMS populations show no apparent variation between different star forming regions, exhibiting shapes consistent with the ONC where the full population is nearly entirely detected with very high sensitivity from the 838~ks observation of the {\it Chandra} Orion Ultradeep Project.  The truncated XLF of a distant region observed with short {\it Chandra} exposures can therefore be calibrated to the ONC's complete XLF to obtain the total stellar population (in units of ONC's population).  This procedure is described and applied to 16 major star forming regions by \citet{Kuhn15}.  

The results for IRAS 09002-4732 are shown in Figure~\ref{xlf.fig}.  The solid histogram shows the distribution of total-band (0.5-8 keV), absorption-corrected X-ray luminosities for 839 lightly-obscured ($\log N_H < 22.0$~cm$^{-2}$) stars with spectral types cooler than B4 from the {\it Chandra} Orion Ultradeep Project \citep{Getman05}.  These are the $\log L_{t,c}$ values for the  the `lightly obscured cool stars' in Table~1 of \citet{Feigelson05}.   The dashed histogram shows the distribution of $L_X$ values for the 386 IRAS 09002-4732 {\it Chandra} stars.   The sample becomes incomplete below $\log L_X \simeq 31.0$ erg~s$^{-1}$, but the region clearly has more X-ray stars with $\log L_X \sim  31-32$ erg~s$^{-1}$ than the ONC.  

A maximum likelihood fit to a power law distribution for $\log L_x \geq 31.0$ erg~s$^{-1}$ with slope 0.9 gives a normalization factor of 2.3; that is, the IRAS 09002-4732 region shown in Figure~\ref{acis_disk.fig} contains more than twice the unobscured population of the ONC.  If the ONC has an estimated population of 2600 stars  down to the stellar limit of 0.08~M$_\odot$ \citep{Kuhn15}, then the IRAS 09002-4732 region has $\sim 6,000$ stars.  If we adopt a relative 80\%-20\% populations of the two ellipsoidal structures discussed in \S\ref{cluster.sec}, then the total population of the main cluster is $\sim 5,000$ stars and the total population of the younger component in the molecular filament is $\sim 1,000$ stars within the {\it Chandra} field of view.

\section{Discussion}\label{discussion.sec}

\subsection{Comparison with Other Rich Young Clusters} \label{ONCcomp.sec}


The central IRAS 09002-4732 cluster resembles other rich MYStIX clusters modeled by Kuhn et al. (2014, 2015).  It shows morphology, population similar to RCW 38, Trumpler 14 in the Carina complex, NGC 6616 in the Eagle Nebula, and the ONC in the Orion complex.  All of these clusters have several thousand total stars and exhibit a core-halo structure where the central isothermal ellipsoidal component has core radius $r_c \simeq 0.1$~pc.  IRAS 09002-4732 differs from the others in three respects:
\begin{enumerate}
\item  The ellipticities of both the core and halo components in IRAS 09002-4732 are higher than in the other clusters where core ellipticities are $\epsilon \lesssim 0.3$ (rather than 0.43) and halo ellipticities are $\epsilon \lesssim 0.5$ (rather than 0.84).  

\item IRAS 09002-4732 is more heavily absorbed than the Carina, Eagle and Orion clusters with $A_V$ reaching $\sim 50$ mag in the cluster core compared to  $A_V < 5$ mag for the other cases.  Only RCW~38 shows similarly high absorption.  

\item Except for RCW~38, IRAS 09002-4732 is younger than the comparison clusters with $Age_{JX} \sim 0.8$ Myr compared to a more typical $Age_{JX} \sim 2$ Myr. Appendix~B presents new evidence that RCW~38 has a very young age similar to IRAS 09002-4732. 
  
\end{enumerate}

IRAS 09002-4732 and RCW 38 thus resemble some of the richest young stellar clusters in the nearby Galactic Plane but seen at an earlier stage of evolution with less time for dynamical equilibration (hence higher eccentricity) and less time for destruction of the molecular environment (hence higher absorption).  This latter effect is also reflected in the size of the ionized H~II region: Orion and Eagle have giant H~II regions $> 1$~pc in extent while IRAS 09002-4732 and RCW~38 have UltraCompact H~II regions only $\simeq 0.1$~pc in extent \citep{Apai05}\footnote{
The small evacuated bubble around the dominant star of RCW~38 is shown by \citet{DeRose09}. 
Trumpler 14 does not produce an H~II region because it is born in a complex where supernova remnants have previously dissipated most of the molecular material \citep{Townsley11a, Townsley11b}.}. 

Table~\ref{ClusComp.tbl} provides a more quantitative comparison of the IRAS 09002-4732 stellar population to the Orion Nebula Cluster region.  As outlined in \S\ref{intro.sec}, the Orion region has motivated much of the astrophysical discussions of cluster formation \citep{Krumholz19, Vazquez19}.   Combining the ellipsoidal structural parameters obtained in  \S\ref{cluster.sec} above with the estimates of total population in \S\ref{pop.sec}, cluster physical parameters can be derived following \citet{Kuhn15, Kuhn15b}.  The quantities compared are: a characteristic radius $r_4 = 4 r_c$ in units of pc, roughly corresponding to the projected half-mass cluster radius; the total stellar population within this radius, $n_4$; the stellar surface density $\Sigma_0$ within this radius in stars~pc$^{-2}$; and the central stellar density $\rho_0$ within this radius in stars~pc$^{-3}$.   In Table~\ref{ClusComp.tbl}, the Orion Nebula `Cluster core' corresponds to the small and dense Trapezium component, the `Cluster halo' corresponds to the main portion of the ONC, the `Protostars' corresponds to the north-south molecular filament pointing towards the OMC-2/3 star forming region.  (A fourth elliptical component corresponding to the Becklin-Neugebauer/Kleinman-Low protostellar cluster in the OMC-1 cloud behind the ONC is omitted here.)  

\begin{deluxetable}{lccccccc}
\tablecaption{Comparison of IRAS 09002-4732 and Orion Nebula Cluster Properties \label{ClusComp.tbl}}
\tabletypesize{\small}
\tablewidth{0pt}
\centering
\tablehead{
\colhead{Cluster} & \colhead{$\log r_4$} & \colhead{$\log n_4$} & \colhead{$\log \Sigma_0$} & \colhead{$\log \rho_0$} & \colhead{$\epsilon$} &  \colhead{ME} & \colhead{$\log$Age}	 \\
 & \colhead{pc} & \colhead{stars} & \colhead{stars pc$^{-2}$} & \colhead{stars pc$^{-3}$} &  \nodata & \colhead{keV} & \colhead{yr} 
}
\startdata
\multicolumn{8}{c}{\bf IRAS 09002-4732} \\
Cluster         &  -0.37 & 3.57 & 4.55 & 5.22 & 0.43  & 2.5-5    & 5.85 \\
Protostars    & ~0.36 & 2.96 & 2.49 & 2.43 & 0.84 & \nodata & 5.0: \\
\multicolumn{8}{c}{\bf Orion Nebula} \\
Cluster core & -0.71 & 2.17 & 3.85 &  4.86 & 0.30  & 1.6 & 6.04 \\
Cluster halo & -0.06 & 3.21 & 3.58 &  3.94 & 0.49  & 1.6 & 6.18 \\
Protostars    & -0.44 &1.94 & 3.07 &  3.81 & 0.84   & 1.4 & 5.0: \\
\enddata
\end{deluxetable}

As with our preview examination, Table~\ref{ClusComp.tbl} shows that the IRAS 09002-4732 cluster is richer, more embedded, and younger than the ONC.  Its core radius is intermediate between the ONC Trapezium and ONC halo components.  The inferred central star density is thus extremely high, $\rho_0 \simeq 1.5 \times 10^5$ stars~pc$^{-3}$.  This is similar to the central density inferred for RCW~38, $\rho_c \simeq 2 \times 10^5$ stars~pc$^{-3}$ \citep{Kuhn15b}.  If the ONC Trapezium experienced close stellar encounters and dynamical ejections, as suggested by \citep{Pflamm06} based on its anomalous mass function, then similar stellar interactions may be present in IRAS 09002-4732 today.  

Both the IRAS 09002-4732 and ONC rich clusters have substantial ellipticity, with the former around $\epsilon \simeq 0.6=0.7$ and the latter around $\epsilon \simeq 0.3-0.5$.  But their associated protostellar populations have even higher ellipticities ($\epsilon \simeq 0.8-0.9$); these spatial distributions can be considered to be `filamentary' \citep{Getman18b}.  And in both cases, the protostars are embedded in filamentary molecular cloud structures.   This contrasts with older rich clusters, such as NGC~6231 \citep{Kuhn17} and NGC~2244 in the Rosette Nebula \citep{Kuhn15b}, that are much larger and nearly spherical.  Gaia proper motions show that these (and other) young clusters are dynamically expanding from an earlier compact state \citep{Kuhn19}. 

The physical properties of the central cluster of IRAS 09002-4732 (\S\ref{properties.sec}) can also be compared to the full sample of MYStIX (sub)clusters by placing it on the bivariate scatter plots shown in Figure 1 of \citet{Kuhn15}.  For some properties, the central cluster follows trends of young age, high absorption, small radius, and high central star density. But it has an unusual combination of other properties.  Trumpler~14 in the Carina Nebula and NGC~6611 in the Eagle Nebula are nearly as rich as IRAS 09002-4732, but they are older and less absorbed with core radii several times larger and consequently much lower central star densities.  In the case of NGC~6611, there is clear evidence of kinematic expansion from a more compact morphology from analysis of $Gaia$ proper motions \citep{Kuhn19}.  The core of RCW~38 is the closest match to the IRAS 09002-4732 central cluster with similar high population,  young age (Appendix B), small radius, and high central star density.  We also note that M~17 is very rich and centrally condensed, but its morphology is `clumpy' without a single component that dominates the population.  

We can also seek analogs to IRAS 09002-4732 based on the early evolutionary stage of its H~II region.  The ultracompact H~II region requires that its most massive member is extremely young and/or just recently encountered the molecular environment.  Most heavily obscured young clusters have relatively large H~II regions such as the less rich W~40 \citep{Rodney08} and the Flame Nebula \citep{Bik03} with $\sim 1$~pc extent of the ionized radio continuum region.  But W3-Main is a rich older cluster with a remarkable assemblage of HII regions ranging from hypercompact ($\lesssim 0.003$~pc) to giant ($\simeq 0.5$~pc) in size \citep{Tieftrunk97, Feigelson08, Bik14}.   Collectively, the evidence suggests that massive star formation can occur over a long timescale but is sometimes delayed until after most lower mass stars have formed and assembled into a cluster.

In summary, seven very rich young stellar clusters are present in the nearby Galactic Plane:  the ONC, Trumpler~14, NGC~6611, M~17, W3~Main, RCW~38, and IRAS 09002-4732.  Some have very dense cores with $\gtrsim 10,000$ stars~pc$^{-3}$ while others more extended, likely expanded from a more compact state.    IRAS 09002-4732 and RCW~38 represent the earliest stage of evolution with the highest absorption and youngest stars.  

\subsection{Implications for star cluster formation} \label{implSF.sec}

First seen in large-scale CO maps of the Orion complex clouds \citep{Bally87, Uchida91}, filamentary molecular structures are now found on many scales throughout the interstellar medium.  The Snake Nebula is an isolated sinuous cloud $\simeq 30$~pc long with a hierarchy of dense cores on scales $0.1-0.01$~pc \citep{Wang14}.  Regions in Aquila Rift are composed of a networks of filaments \citep{Moriguchi01, Kusune16}.   The emission line nebula of IRAS 09002-4732 is partly obscured by dark filaments with characteristic lengths around 0.5~pc and widths around 0.05~pc with scattered small dark globules within \citep[][see Figure~\ref{acisJHK.fig} below, lower right panel]{Apai05}. Typically some filament concentrations are undergoing gravitational collapse with protostars within while other concentrations are currently stable.   

Hydrodynamical models of star formation in realistically turbulent molecular cloud cores show that filaments form quickly, undergoing fragmentation into cores that form small groups of stars \citep{Bate03, Federrath15, Vazquez17, Vazquez19}.   These groups are dragged into the center of the cloud gravitational potential well as the gaseous filaments infall and merge. In early calculations lasting $\lesssim 1$~Myr, the stars of different ages are efficiently mixed into a homogeneous, quasi-spherical cluster (Bate et al.).  But our empirical research shows that rich clusters show a strong spatial age gradient on timescales of $\sim <0.5$ to 5~Myr, requiring that star formation continue in the cluster cores after cluster formation \citep{Getman18}.  Some later calculations can explain this effect through a hierarchical filament merging process, combined with stellar feedback, continuing over millions of years (Vazquez-Semadeni et al.).    

Discussion continues over the relative importance of turbulence and thermal pressure, magnetic dynamical effects, low-mass and high-mass stellar feedback onto the collapsing gas, binary star systems and N-body interactions in the denser cluster cores \citep{Krumholz19}. Turbulence may play a more limited role than originally thought. But the basic scenario of slow, hierarchical cluster formation via merging filaments seems well-established from an astrophysical perspective, in contrast to older models involving rapid monolithic cluster formation \citep{Elmegreen00} or a global exponential burst of star formation \citep{Stahler00}.  

However, there is little {\it direct} empirical evidence that demonstrates that {\it rich} clusters (containing thousands of stars) form with molecular filaments.  The ONC is an example of not-infrequent {\it indirect} associations of an older rich clusters with an origin in filaments.   The link is indirect because the molecular material giving rise to the ONC has been destroyed by the giant H~II region.  First, many sparse groups of protostars are forming today in the contemporary Orion molecular filament \citep{Megeath12}, although we do not know whether they will consolidate into a rich cluster.  Second,  the ONC has an elliptical morphology elongated long the filament axis.  Third, stars projected against the OMC-1 filament show radial velocities similar to that of the OMC-1 gas; this star-cloud velocity agreement is evident despite a strong velocity gradient along the filament \citep{Getman19}.  

IRAS 09002-4732 is a rare case where a nascent rich star cluster, possessing a massive O star and thousands of PMS stars, is embedded in a molecular filament.  \citet{Apai05} analyze the three-dimensional environment and conclude that the cluster lies between dense cores; that is, the cluster is truly embedded and not just obscured behind a screen of cloud material.  The H~II region of the dominant O7 star is in the ultracompact stage suggesting that the star is very recently formed, perhaps in the past $10^4$ years.  While the average age of the cluster stars is $\leq 0.8$~Myr (\S\ref{age.sec}),  an age gradient is quite possibly present \citep{Getman18} and the central core may be actively forming stars today.  The surrounding natal molecular environment is in process of being heated and ejected, and the cluster core is so dense (\S\ref{ONCcomp.sec}) that high-velocity N-body interactions (or even stellar collisions) should be occurring.  This recent and rapidly evolving activity takes place within a molecular filament several parsecs in extent that exhibits mild, distributed star formation outside of the central cluster.  

\subsection{Initial Mass Function of IRAS 09002-4732} \label{IMF.sec}

Finally, we note that the IRAS 09002-4732 cluster may have an anomalous Initial Mass Function (IMF) in the sense that it is deficient in massive stars.  If it truly has a stellar population $2-3$ times that of the ONC, which itself may lack a full complement of O stars, then it is expected to have more massive stars.  A standard IMF with a Salpeter power law tail is expected to have $\simeq$20 stars above $M =10$~M$_\odot$ and $2-3$ stars above $M = 30$~M$_\odot$ for a population around $5000-8000$ stars (\S\ref{pop.sec}) above the stellar limit \citep{Kroupa01}.  We would thus expect to see a collection of small H~II regions in the cluster core rather than just one around Source D. 

If this deficiency of massive stars is real, two explanations are plausible.  First, the stars may have formed but are now ejected by dynamical interactions with other massive stars.  Rapid dynamical decay has been calculated to be responsible for the deficit of massive stars in the ONC core \citep{Pflamm06, Allen17} which we estimate is less dense than the core of IRAS 09002-4732 (Table~\ref{ClusComp.tbl}).  This hypothesis can be empirically tested by a search for stars recently ejected from the cluster core in the Gaia proper motion database.    Second, it is possible that massive stars other than Source D have not yet formed.  There is evidence for delayed O star formation in the ONC \citep{ODell09},  W3 Main \citep{Tieftrunk97, Feigelson08, Bik14}, and M~17~SW \citep{Povich16}.   This picture is consistent with the global hierarchical gravitational collapse by \citet{Vazquez19} who explain late massive star formation in the central potential well and later destruction of filamentary gas feeding by OB stellar feedback.

\section{Conclusions}

We obtain a new census of PMS stars for the IRAS~09002-4732 star forming region from a deep observation with the {\it Chandra} X-ray Observatory. In concert with existing optical and infrared catalogs, 386 probable members with sub-arcsecond positions are identified (Table~\ref{acis_mem.tbl}), both within the central cluster and in the surrounded cloud.  These are supplemented by 55 non-X-ray infrared-excess stars (Table~\ref{MIRES.tbl}).  The sample selection, following procedures from our MYStIX and SFiNCs surveys, has low contamination by Galactic field stars or quasars.   Measures of magnetic activity ($\log L_x$ obtained from X-ray photometry), absorption ($\log N_H$ and $A_V$ from X-ray median energy), age ($Age_{JX}$ from X-ray and near-infrared photometry, and optical ages from an $r-i$ color-magnitude diagram), and disk presence (from near- and mid-infrared photometry) are calculated for the 441 individual stars.  

From this information, collective properties of the cluster and star forming environment are inferred: 
\begin{enumerate}
\item  The stars have a distinctive bipartite spatial structure.  Most members reside in a compact, elliptical  rich cluster with core radius 0.1~pc centered 0.15~pc north of Source D, the O7 star that dominates the light from the cluster.  A minority reside in a loose, highly elongated (eccentricity $\sim 90$\%) distribution oriented northwest-southeast around the cluster.  This coincides with a molecular filament several parsecs long that connects to the nearby, and larger, molecular cloud hosting the rich RCW 38 young star cluster. (\S\ref{filament.sec}, \ref{cluster.sec})
 
\item The distance to the member stellar population is approximately 1.7~kpc, evaluated from Gaia parallax measurements.  This is consistent with most previous distance estimates. (\S\ref{dist.sec})

\item The stellar population is heavily absorbed with median absorption $A_V \simeq 13$ mag. The extinction increases sharply to $A_V \simeq 50$ mag at the cluster center.  (\S\ref{extinct.sec}) 

 \item The average age of the member population, both from X-ray/infrared photometry ($Age_{JX}$) and from an $r-i$ color-magnitude diagram, is $\leq 0.8$~Myr calibrated to the evolutionary tracks of \citet{Siess00}.    But the filament-associated elongated population is younger, as it is identified by several dozen protostars, and a spatial gradients in age within the main cluster may also be present.  Source D may also be very young as its H~II region is very small (ultracompact) with 0.1~pc extent.  (\S\S\ref{age.sec}, \ref{ONCcomp.sec})
 
 \item The stellar population is remarkably large. The total population deduced from the X-ray luminosity function is twice that of the Orion Nebula Cluster, or roughly 5000 stars. (\S\ref{pop.sec}) 
 
 \item The cluster core is incredibly dense with $\sim 1.5 \times 10^5 $ stars~pc$^{-3}$.  This is a regime where strong N-body interactions should be common causing destruction of binaries, high-velocity ejection of runaway members, and possibly stellar collisions. (\S\S\ref{ONCcomp.sec}, \ref{IMF.sec})
 
 \item In morphology and population, the IRAS 09002-4732 cluster resembles some of the richest young clusters in the solar neighborhood: Orion Nebula Cluster in the Orion Nebula, Trumpler 14 in Carina, NGC 6611 in Eagle, M~17 in Omega, W3 Main in W~3,  and the nearby RCW~38.  But IRAS 09002-4732 and RCW~38 are distinctively younger, more heavily absorbed (with less time for stellar feedback to disperse the gas), with higher ellipticity (with less time for dynamical equilibration), and smaller H~II regions. (\S\ref{ONCcomp.sec})

\item The Initial Mass Function of the cluster appears to be deficient in massive stars.  Either OB stars similar to Source D have not yet formed, or have been dynamically ejected by N-body interactions in the dense core. (\S\ref{IMF.sec})

 \end{enumerate}

 We conclude that IRAS 09002-4732 is a rare case where a rich (thousands of stars) cluster is forming today in a molecular filament.  More commonly, either the cluster is sparse (dozens of stars) or the rich cluster has already destroyed its natal molecular environment.  The stellar spatial distribution, individual properties (ages, extinction) and collective properties (population, IMF, cluster structure)  directly support contemporary numerical calculations of cluster formation in clouds that involve the hierarchical formation and merging of groups in molecular filaments such as those described by \citet{Vazquez19} and \citet{Krumholz19}.  
 
The IRAS 09002-4732 cluster, particularly the core with radius 1.5\arcmin\/ centered at $(\alpha, \delta) = (135.47435^\circ, -47.72820^\circ)$,  thus warrants future study as an important laboratory for the study of cluster formation in molecular filaments.   The most crucial needed observations are  high quality molecular maps obtained with the ASTE and ALMA telescopes that might show gravitational infall of gas, either along or perpendicular to the filament axis.   In the immediate proximity of the cluster, combinations of gas infall to fuel star formation and gas outflow in response to stellar feedback may be occurring.  The complicated gas kinematics that may be present could benefit from study with advanced instruments like the optical band MUSE integral field spectrograph on the Very Large Telescope, or the NIRSpec spectro-imager on board the James Webb Space Telescope.    For the stellar population, the Gaia catalog can be searched for high-velocity stars ejected from the core, and infrared multi-object spectroscopy of cluster members can elucidate its dynamical state. Deeper {\it Chandra} observations would identify a larger fraction of the PMS stellar population both in the cluster and the surrounding filamentary cloud.    We also recommend further study of the gas and stellar content of W3~Main, another rare case of an embedded rich star cluster where the diversity of H~II regions points to continuing star formation in the cluster core \citep{Feigelson08}. 

\acknowledgements

{\it Acknowledgements} The first two authors contributed equally to this study.  We benefited from valuable discussion with Daniel Apai (Arizona) and insightful comments of an anonymous referee.   This work was supported by NASA grants GO7-18001X and the {\it Chandra} ACIS Team contract SV474018 (G. Garmire and L. Townsley, Principal Investigators), issued by the {\it Chandra} X-ray Center, which is operated by the Smithsonian Astrophysical Observatory for and on behalf of NASA under contract NAS8-03060. The Guaranteed Time Observations (GTO) data used here were selected by the ACIS Instrument Principal Investigator, Gordon P. Garmire, of the Huntingdon Institute for X-ray Astronomy, LLC, which is under contract to the Smithsonian Astrophysical Observatory; Contract SV2-82024. 

This work has also made use of data from the European Space Agency (ESA) mission {\it Gaia} processed by the {\it Gaia} Data Processing and Analysis Consortium (DPAC). Funding for the DPAC has been provided by national institutions, in particular the institutions participating in the {\it Gaia} Multilateral Agreement.  This publication makes use of data products from the Two Micron All Sky Survey, which is a joint project of the University of Massachusetts and the Infrared Processing and Analysis Center/California Institute of Technology, funded by the National Aeronautics and Space Administration and the National Science Foundation.  This research has made use of NASA's Astrophysics Data System Bibliographic Services and SAOImage DS9 software developed by Smithsonian Astrophysical Observatory.

\appendix
\section{Core of IRAS 09002-4732} \label{core.sec}

While it had long been established that IRAS 09002-4732 was dominated by an O7 star with high luminosity, $L \simeq 1 \times 10^5$~M$_\odot$ \citep{Lenzen91}, high-resolution infrared imaging with ESO's Very Large Telescope demonstrates that the core has a dense concentration of OB stars \citep{Apai05} that is surrounded by a cluster of lower mass PMS stars.  Our {\it Chandra} ACIS images have similar resolution and detect a considerable sample of cluster members.  This provides a motivation to compare the {\it Chandra} and VLT images.

The dominant star in the cluster is located at $09^h 01^m 54.45^s -47^\circ 44^\prime 11.0^{\prime\prime}$ (J2000) with designations IRAS 09002-4732, IRS8 \citep{Lenzen91}, S13 \citep{Ghosh00}, Source D \citep{Apai05}, and 2MASS J090154447-4744112. We will call this `Source D' following \citep{Apai05}, noting that SIMBAD database does not make all of these associations.  The 2MASS star is bright in the 2~$\mu$m band with $K=11.4$ but is heavily reddened with $J-K>5.5$ ($A_V>11$ mag).  With flux density $\sim$ 15,000 Jy at 100~$\mu$m, it is one of the most luminous infrared sources of the Vela molecular clouds.  For comparison, the infrared-brightest protostar in the nearby molecular filament is sequence \#386 associated with MCP RCW38 69 =  2MASS J09020115-4744563 = MSX6C G268.4452-00.8442 = [MHL2007] G268.4452-00.8442 1 \citep{Romine16}. At 10.4~$\mu$m, the dominant cluster star has flux density 38.3~Jy while the brightest protostar has 0.27~Jy \citep{Mottram07}.  \label{note1}

\begin{figure}[h]
\centering
\includegraphics[width=0.46\textwidth]{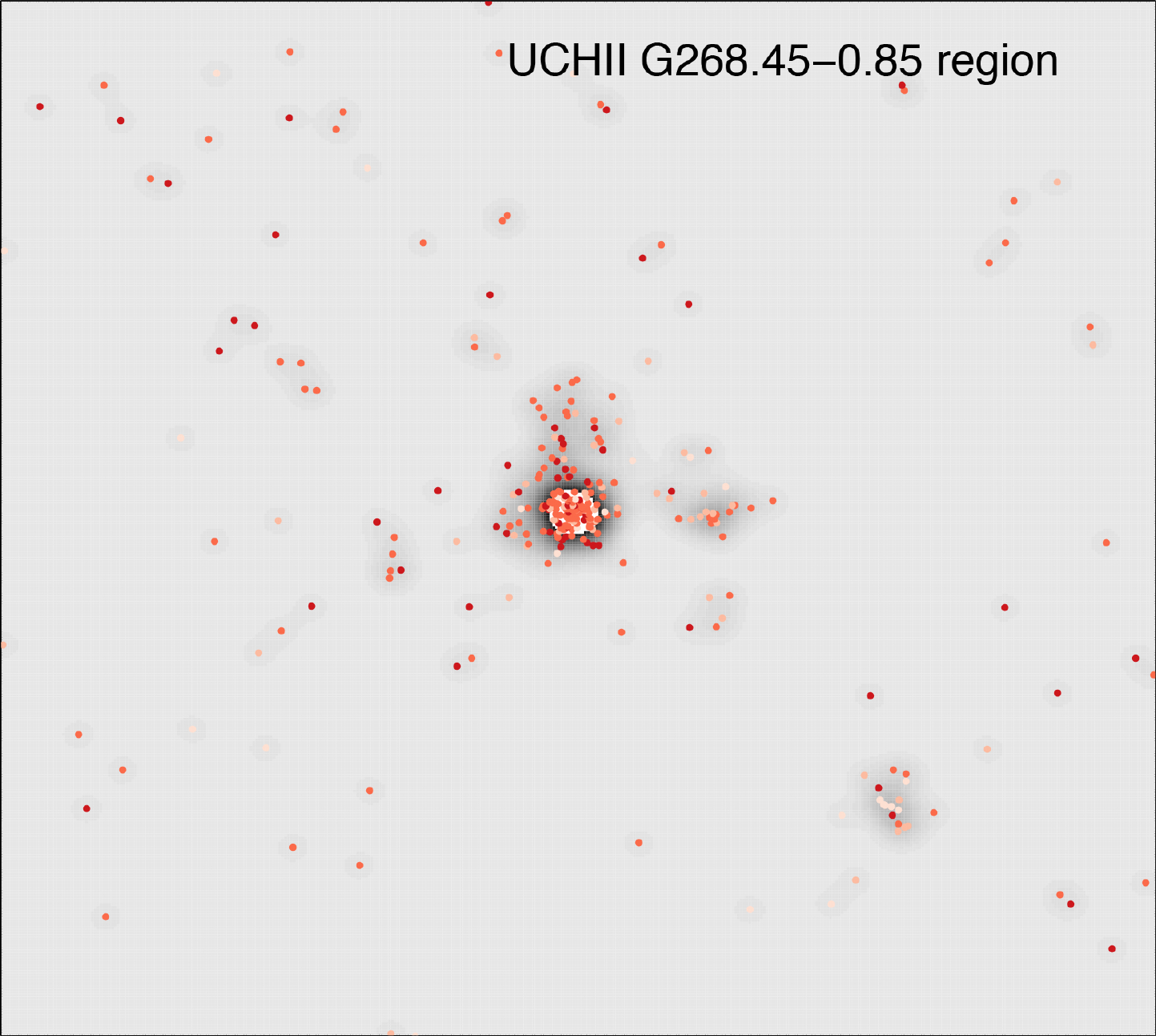} \hspace{10pt}
\includegraphics[width=0.42\textwidth]{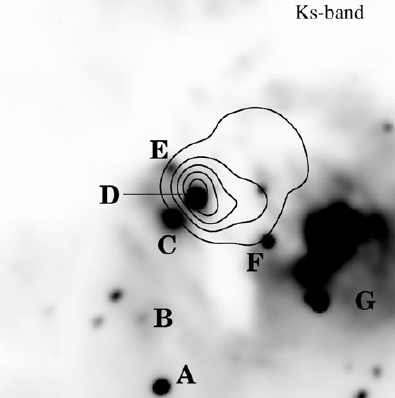} \\ \vspace{10pt}
\includegraphics[width=0.46\textwidth]{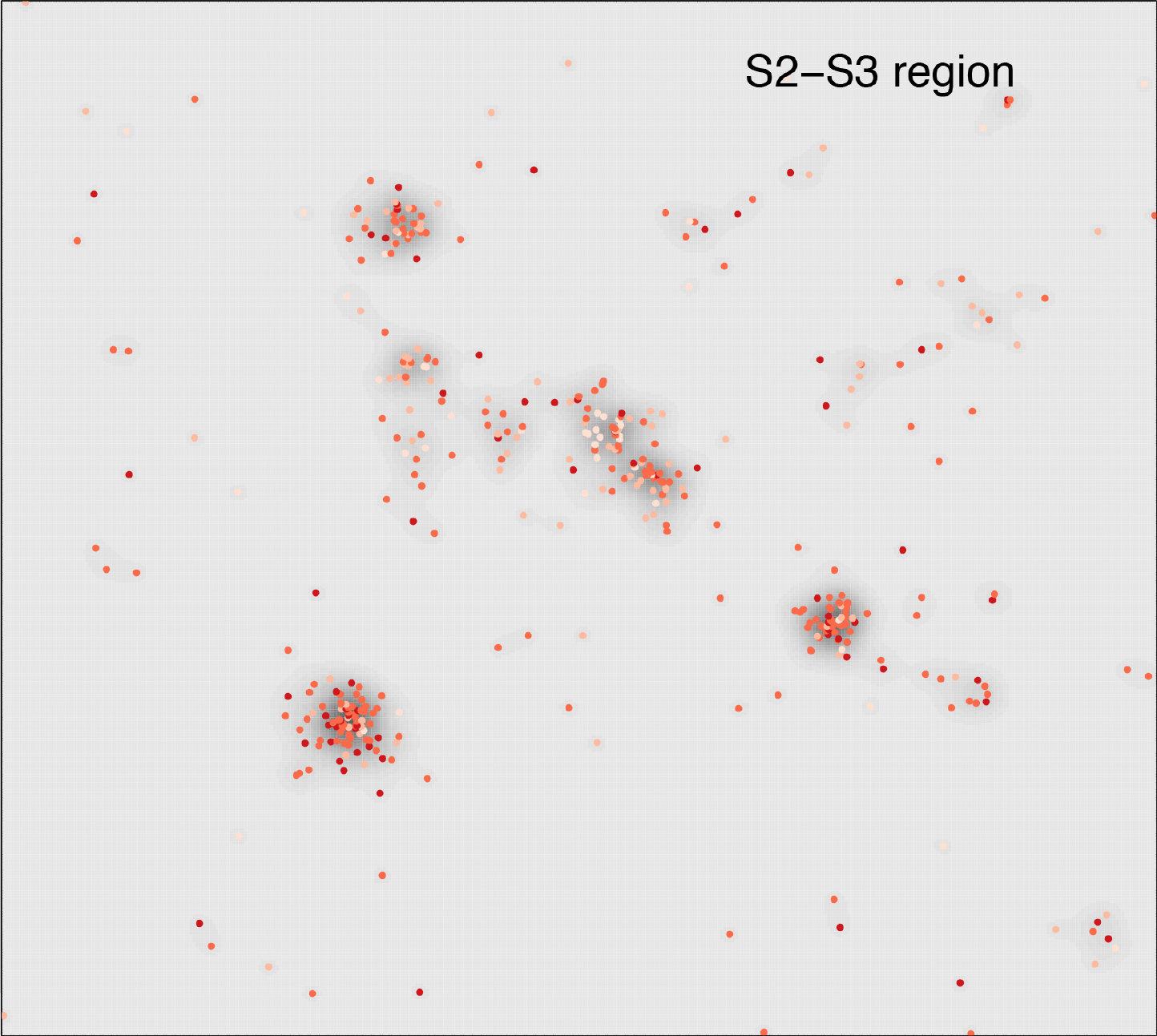} \hspace{10pt}
\includegraphics[width=0.42\textwidth]{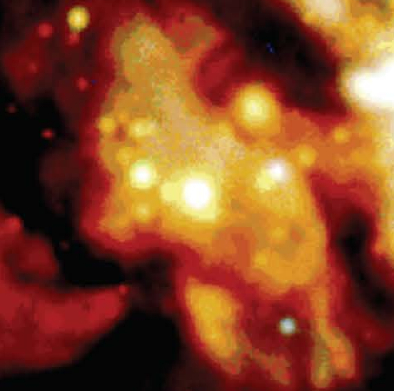} 
\caption{Comparison of X-ray and infrared images for two small regions in the core of IRAS 09002-4732.  Each image subtends $20\arcsec \times 20\arcsec$ {\it Top:} Source~D and the ultracompact H~II region G268.45-0.85, {\it Chandra} ACIS and VLT/ISAAC $K_s$ band \citep{Apai05}. {\it Bottom:} A low-obscuration window between dust filaments S2 and S3; {\it Chandra} ACIS and a VLT/ISAAC $JHK$ composite image.  The ACIS image shows individual photons with redness indicating hardness (typical energies $2-7$~keV) superposed on a gray-scale map smoothed with a $\sigma=1\arcsec$ Gaussian kernel.  The upper-right panel shows radio continuum contours delineating the ultracompact H~II region.  The lower-right panel shows the $S2$ and $S3$ sinuous dusty filaments.   \label{acisJHK.fig}}
\end{figure}

Figure~\ref{acisJHK.fig} illustrates the complicated relationship between the X-ray and infrared samples in two small portions of the IRAS 09002-4732 cluster: a 20\arcsec\/ region around Source D with its radio ultracompact H~II region (top panels); and a 20\arcsec\/ region of lower obscuration between the dust filaments $S2$ and $S3$ discussed by \citet[][bottom panels]{Apai05}.  Even with a 100~ks exposure, the {\it Chandra} image is photon-limited; sources with $\simeq 5$ photons are confidently detected, while those with $\sim 2-4$ photons are possible sources.  We also recall that the {\it Chandra} exposure map shows small-scale variations in exposure times due to satellite dithering motions of the CCD chip gaps. 

Source D is not detected in the X-ray band.  The nearest sources are Seq \#190 (090154.33-474410.3, $\log L_x \simeq 31.9$ erg~s$^{-1}$) and \#202 (090154.63-474411.1,  $\log L_x \simeq 31.2$ erg~s$^{-1}$) that lie 1.5\arcsec\/  NW and 1.9\arcsec\/ E of the infrared position of Source D, respectively  (Tables~\ref{candx.tbl}-\ref{acis_mem.tbl}).  These offsets are too large to be attributed to X-ray positional uncertainties ($<0.2$\arcsec) or X-ray/infrared coordinate frame offsets ($<0.1$\arcsec).  There is precedence in the Orion region for the X-ray detection of PMS secondary companions but a failure to detect the massive primary.  These include the Becklin-Neugebauer Object, where two X-ray stars are detected within a few hundred AU of the luminous high-velocity massive star \citep{Grosso05}, and $\theta^1$ Ori B where the B2.5 primary $\theta^1$ Ori B East would be undetected in IRAS 09002-4732 ($\log L_x=30.1$ erg s$^{-1}$ with soft X-ray spectrum that can not penetrate high extinction) but the flaring hard X-ray emission produced by binary low-mass companions $\theta^1$ Ori B West would be detected \citep[$\log L_x=31.5$ erg s$^{-1}$,][]{Stelzer05}. 

Both the infrared and X-ray images show several stars within $\simeq 10$\arcsec\/ of Source D, but the correspondence is not simple.  Infrared stars F and G appear as very faint X-ray sources, as well as an anonymous star 3\arcsec\/ west of Source D.  The more prominent infrared stars A, B, C and E are not seen in X-rays.  The {\it Chandra} image shows barely-resolved stars $1\arcsec-2$\arcsec\/ east and north of Source D that are not evident in the available infrared images.  

The $S2-S3$ region shown in Figure~\ref{acisJHK.fig} (bottom panels) also shows both similarities and differences in the two bands.  As with Source D above, lower mass members of the region are seen in X-rays while the most luminous infrared member is not detected. The X-ray image shows $\sim 9$ faint X-ray sources in a chain extending $\sim$15\arcsec\/ from northeast to southwest.  Most of these have discernible counterparts in the VLT infrared image.  The brightest X-ray source to the southeast lies embedded in or behind a dark dust lane and is thus missing an infrared counterpart.  This source is sequence \#227 (090155.50-474355.7) with 46 net counts, absorption-corrected $\log L_x \simeq 31.6$ erg~s$^{-1}$, and median energy absorption $ME \simeq 3.9$~keV corresponding to $\log N_H \simeq 9 \times 10^{22}$ cm$^{-2}$ and $A_V \simeq 45$~mag.  This star lies near the top of the X-ray luminosity function (Figure~\ref{xlf.fig}).

\section{Age of RCW~38} \label{rcw38_age.sec}   

\citet{Getman14} reported that RCW~38, W~40, and Flame Nebula are the youngest MYStIX star forming regions with average ages of $Age_{JX} \leq 1$~Myr (see their Figure~1), but \citet{Kuhn15} incorrectly reported an older age similar to most MYStIX clusters.  At that time, only 21 MYStIX young stars in RCW~38 had available $Age_{JX}$ estimates and, due to such a poor sampling, RCW~38 was omitted from further detailed age analyses of Getman et al. 

\begin{figure}[h]
	\centering
	\includegraphics[width=0.45\textwidth]{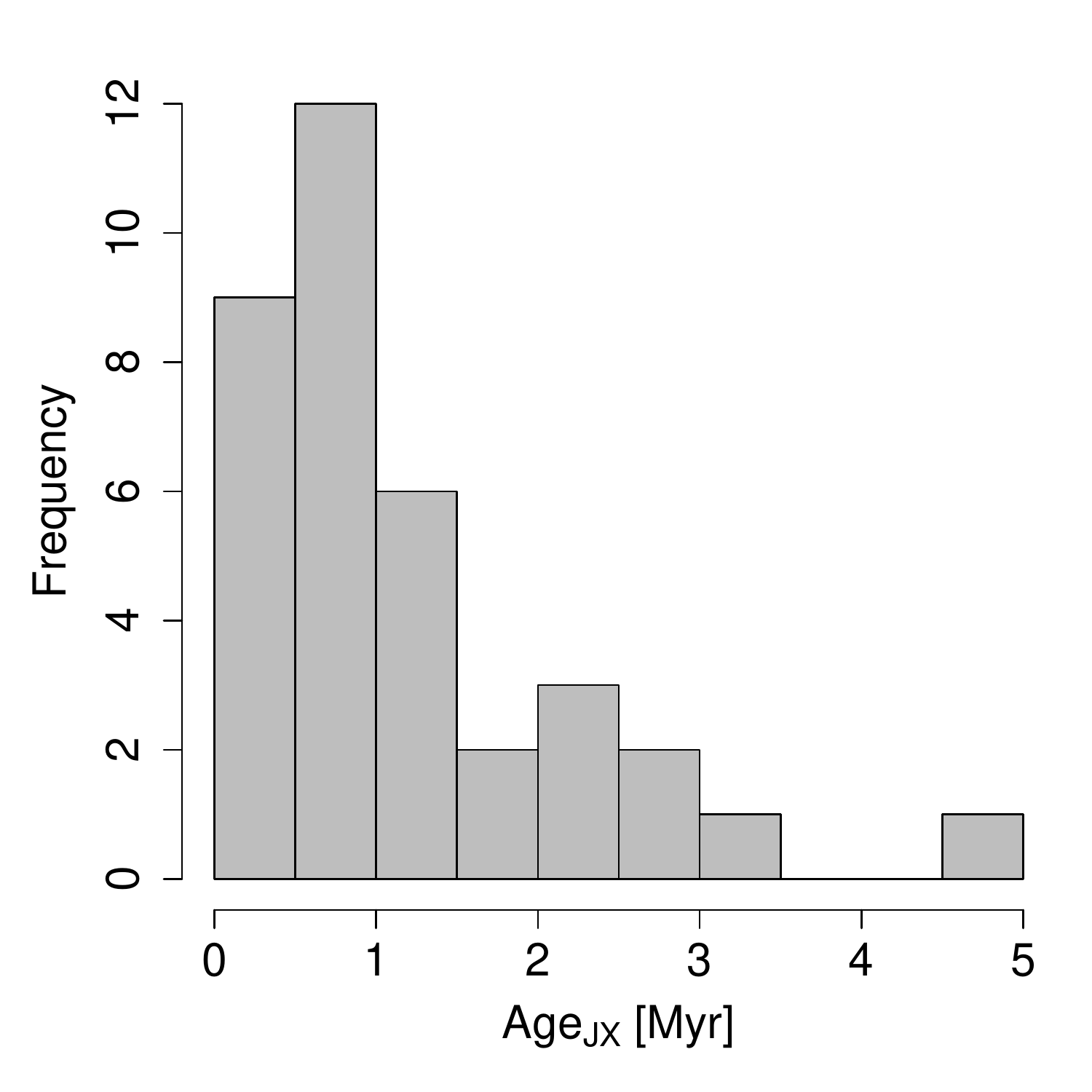}
	\caption{Histogram of stellar ages for the 37 young X-ray stars in RCW~38.  \label{rcw38_ages.fig}}
\end{figure}

We can now clarify this situation by combining the archive $\sim 100$~ks {\it Chandra} exposure of RCW~38 (obsid 2556) used in MYStIX with our new unpublished {\it Chandra} RCW~38 observations (obsids 16657, 17554, and 17681) with similar total exposure.  Together, these observations yield a larger $Age_{JX}$ sample of 37 young stars with $>7-10$ X-ray net counts. Since the near-IR photometry is prone to the effects of high absorption and diffuse nebular emission in the cluster core, all but one of the $Age_{JX}$ stars are located in the halo of the RCW~38 cluster.  We thus are unable to evaluate whether a spatial age gradient is present \citep{Getman18}. The resulting stellar age estimates are listed in Table~\ref{rcw38_ages.tbl} and a histogram of the ages is presented in Figure~\ref{rcw38_ages.fig}. The median age for these 37 stars is $0.9 \pm 0.1$~Myr, and it is likely that the stars in the core are even younger.  We conclude that the cluster RCW~38 has an age close to that of the IRAS 09002-4732 cluster, and both are unusually young among rich clusters in star forming regions in the nearby Galactic Plane.

\begin{deluxetable}{ccc}
	\centering
	\tabletypesize{\tiny} 
	\tablecaption{{\bf $Age_{JX}$ stars in RCW~38}  \label{rcw38_ages.tbl}}
	\tablehead{
		\colhead{CXOU J} & \colhead{$C_{net}$} & \colhead{$Age_{JX}$} \\
		& \colhead{counts} & \colhead{Myr}
	}
	\startdata
	085820.49-473527.3 & 13.9 & 0.5\\
	085821.18-473107.6 & 45.4 & 2.5\\
	085831.44-472807.7 & 49.7 & 1.7\\
	085834.11-472404.6 & 22.7 & 2.5\\
	085835.29-472900.1 & 25.5 & 0.8\\
	085840.64-473430.4 & 14.3 & 0.2\\
	085841.56-472834.6 & 47.4 & 1.5\\
	085849.32-473507.5 &  7.6 & 0.4\\
	085850.80-473012.3 & 15.0 & 1.0\\
	085853.52-472525.2 & 30.7 & 0.9\\
	085853.96-472619.6 & 68.1 & 0.9\\
	085854.97-472803.1 & 11.5 & 0.7\\
	085857.56-472426.0 & 76.4 & 0.7\\
	085859.82-473222.5 & 23.3 & 0.1\\
	085859.89-473319.9 & 13.1 & 0.5\\
	085859.96-473552.5 &  7.2 & 0.6\\
	085859.98-472244.8 & 24.0 & 0.6\\
	085901.48-473003.0 &  8.4 & 0.5\\
	085903.76-472759.5 & 50.0 & 3.0\\
	085910.52-472349.4 & 33.4 & 2.6\\
	085923.68-473834.0 & 14.2 & 4.6\\
	085924.71-473802.6 & 43.2 & 2.3\\
	085924.98-473213.1 & 18.2 & 1.3\\
	085927.72-473551.3 & 26.5 & 1.5\\
	085927.86-472802.9 & 34.4 & 0.8\\
	085931.28-473330.3 & 62.7 & 0.8\\
	085935.41-473828.3 & 10.7 & 1.5\\
	085935.52-473047.9 & 23.1 & 1.0\\
	085937.73-473357.1 & 16.7 & 0.9\\
	085940.23-473100.7 & 11.9 & 0.2\\
	085942.68-472628.4 & 10.5 & 1.1\\
	085944.48-473317.8 & 22.6 & 1.2\\
	085948.39-472640.3 & 41.4 & 0.1\\
	085949.65-473117.1 & 25.5 & 3.2\\
	085957.85-473214.6 & 20.5 & 0.2\\
	090006.73-473032.1 & 34.5 & 1.6\\
	\enddata
	\tablecomments{Column 1: IAU designation. Column 2: X-ray net counts. Column 3: $Age_{JX}$ estimates.}
\end{deluxetable}

\end{document}